\documentclass[aps,prb,twocolumn,superscriptaddress,10pt,reprint]{revtex4-2}

\usepackage{graphicx}
\usepackage{amssymb}
\usepackage{mathrsfs}
\usepackage{amsmath}
\usepackage{amsthm}
\usepackage{braket}
\usepackage{hyperref}
\usepackage{xspace}
\usepackage{booktabs}
\usepackage{multirow}

\usepackage{float}
\makeatletter
\let\newfloat\newfloat@ltx
\let\float@end\float@end@ltx
\makeatother
\usepackage{algorithm}
\makeatletter
\renewcommand*{\fnum@algorithm}{\fname@algorithm~\thealgorithm}
\makeatother
\usepackage{algpseudocode}
\newtheorem{theorem}{Theorem}

\newtheorem{corollary}{Corollary}

\newtheorem{definition}{Definition}

\newcommand{\be}{\begin{equation}}
\newcommand{\ee}{\end{equation}}
\newcommand{\mc}{\mathcal}
\newcommand{\mct}{\textsf{Max-Cut}\xspace}
\newcommand{\prp}{\textsf{PRP}\xspace}
\newcommand{\np}{\textsf{NP}\xspace}
\newcommand{\rt}{\textsf{Rank-2}\xspace}
\newcommand{\gw}{\textsf{GW}\xspace}
\newcommand{\JJ}{\mbox{$J_1$--$J_2$}\xspace}

\begin{document}

\newcommand*{\UA}{Department of Physics and Astronomy, University of Alabama, Tuscaloosa, 35487, Alabama, USA}\affiliation{\UA}

\title{Graph-Theoretic Formulation of Hamiltonian and Ground-State Sign Structures in Heisenberg Antiferromagnets}

\author{Mahmud Ashraf Shamim}
\email{mashamim@crimson.ua.edu}
\affiliation{\UA}
\author{Md Moshiur Rahman Raj}
\email{moshiur\_rahman@tutanota.com}
\affiliation{Department of Physics, University of Rajshahi, P.O. Box 6205, Rajshahi, Bangladesh}
\author{Mohamed~Hibat-Allah}
\affiliation{Department of Applied Mathematics, University of Waterloo, Ontario
Canada N2L 3G1}
\affiliation{Vector Institute,  Toronto,  Ontario,  M5G 0C6,  Canada}
\author{Ruy Fabila-Monroy}
\affiliation{Departamento de Matemáticas, Cinvestav, Mexico}
\author{Nobuchika Okada}
\affiliation{\UA}
\author{Paulo T. Araujo}
\affiliation{\UA}
\date{\today}

\begin{abstract} 
We study Heisenberg antiferromagnets through a graph-theoretic representation of their Hamiltonians. In the computational basis of a fixed-magnetization sector, the Heisenberg Hamiltonian equals a constant minus the weighted Laplacian of a graph whose vertices are spin configurations and whose edges connect configurations related by exchanging two antiparallel spins on a lattice bond. We call this graph the Hilbert graph (HG) which is isomorphic to the token graph of the lattice over which the Hamiltonian is defined. The exact ground-state energy is then determined by the largest eigenvalue of the HG Laplacian, which is the spectral relaxation of weighted \mct on the HG. With the amplitudes fixed, the phase-dependent part of the variational energy defines an antiferromagnetic classical XY model on the HG, with couplings set by the amplitudes. Restricting the variational state to be real, i.e., restricting its phases to $\{0,\pi\}$, reduces the fixed-amplitude energy minimization to solving an antiferromagnetic Ising model and hence a weighted \mct on the HG. When the weights are generated from non-degenerate ground-state amplitudes, the optimal phases give the ground-state sign structure which can be recovered from the weights in time polynomial in the size of the HG. On the Hamiltonian side, a diagonal sign transformation removes all positive off-diagonal matrix elements if and only if the HG is bipartite, which reproduces the Marshall sign rule on bipartite lattices through the graph two-coloring problem; otherwise, the minimal remaining positive weight equals the sum of HG edge weights left uncut by the \mct. This graph formulation establishes a common framework for understanding wave function sign structure, Hamiltonian stoquasticity, non-convexity of phase optimization in variational Monte Carlo, and the average sign in quantum Monte Carlo.
\end{abstract}

\maketitle


\section{Introduction}\label{sec:intro}

Geometrically frustrated Heisenberg antiferromagnets (HAFs) remain a central challenge in quantum magnetism. Frustration prevents all exchange interactions from being satisfied simultaneously, which produces strong competition among low-energy states. Exact ground states are known only in a few special cases, most notably the dimer-product state at the Majumdar--Ghosh point~\cite{Majumdar1969I, Majumdar1969II} and the dimer phase of the Shastry--Sutherland model~\cite{Shastry1981}. Consequently, our understanding of these systems has largely relied on variational methods for ground state (GS) computation, including the density-matrix renormalization group~\cite{White1992,Schollwock2011}, tensor networks~\cite{Orus2014}, and variational Monte Carlo (VMC)~\cite{sorella}, and the difficulty of studying frustrated HAFs is closely tied to the limitations of these methods.

In VMC, a wavefunction ansatz is optimized using energy and gradient estimates from configurations sampled according to its Born distribution~\cite{sorella}. If amplitudes and phases are both trainable, VMC learns them together to approximate the GS. These ansätze broadly fall into three classes~\cite{lange2024architectures}: physics-based forms such as Gutzwiller-projected states~\cite{Gutzwiller1965}, Jastrow-correlated states~\cite{Jastrow1955}, and Pfaffian wavefunctions~\cite{Moore1991}; neural quantum states (NQS)~\cite{Carleo2017}, in which a neural network represents the wavefunction; and hybrids of the two~\cite{Nomura:2017tgw, Ferrari2019}. Among the three classes, NQS are distinguished by universal approximation~\cite{Cybenko1989, Hornik1991}: a sufficiently large network can represent any function, and corresponding expressibility results have been established for many-body quantum states~\cite{Deng2017, Gao2017}.  NQS can therefore represent both the amplitudes and signs of the GS wavefunction in principle, but universal approximation theorems do not let us predetermine the network size, nor do they guarantee accurate sign learning through variational optimization. In frustrated HAFs, it has been observed that learning the signs is considerably harder than learning the amplitudes~\cite{Westerhout:2019hvt,Bukov,szabo}. In practice, accuracy often improves for various NQS architectures~\cite{RNN1, RNN2, RNN3,CNN1, CNN2, CNN3, CNN4,Shamim:2025rsc}, when the sign structure is supplied explicitly as a sign prior, most commonly the Marshall sign rule (MSR)~\cite{marshall1955}; in hybrid ansätze, the prior is supplied by a pair-product factor and optimized alongside the amplitudes~\cite{Nomura:2017tgw, Ferrari2019}.

Many of these architectures have been benchmarked on the square lattice $J_1$--$J_2$ HAF. At $J_2=0$, the MSR gives the exact GS sign structure, and it remains a good approximation for small $J_2$~\cite{Richter1994}. In the strongly frustrated window $J_2/J_1\approx0.5$ to $0.6$, however, no reliable sign prior is known, and performance at $J_2/J_1=0.5$ has become the benchmark by which NQS architectures are judged. Table~I of Ref.~\cite{ViT3} compares variational energies on the $10\times10$ square lattice at this point: most of the ansätze listed there use the MSR, while the Deep Vision Transformer~\cite{ViT3} and group convolutional network~\cite{roth2023high} reach competitive energies without it, as do several architectures on the triangular and other frustrated lattices~\cite{ViT1,ViT2,ViT4,viteritti2026approaching,Kochkov2021}. Architectural design can therefore ease the learning of the GS sign structure, but it does not explain what makes that structure hard to learn.

Because NQS training is largely a black box, understanding the origin of
this difficulty is important for further progress on frustrated HAFs. A few studies have tried to explain this through \textit{complexity}. Westerhout et al.~\cite{Westerhout:2022lil} fixed the GS magnitudes at their exact-diagonalization values and mapped sign reconstruction onto an auxiliary classical Ising model. For the frustrated HAFs they studied, the induced Ising models had \textit{non-glassy} optimization landscapes, and a greedy algorithm recovered the signs accurately. Schurov et al.~\cite{Schurov:2025hcm} analyzed the Boolean Fourier spectra of GS sign structures and showed that several spectral features correlate with learning difficulty. Both studies suggest that sign-learning difficulty is encoded in an identifiable structure of the GS signs; what determines this structure,
however, has remained unaddressed. 

In this work, we study the HAF in a graph-theoretic framework.
In a fixed-magnetization sector, the computational-basis configurations
form the vertices of a \textit{weighted} graph in which two configurations are joined by an edge whenever the Hamiltonian has a nonzero off-diagonal element between them, with that element as the edge weight. Westerhout \textit{et al}.~\cite{Westerhout:2022lil} defined their auxiliary Ising model on this graph, restricted to a subset of the basis. We call it the Hilbert graph (HG), since its vertices form a basis of the sector Hilbert space. We show that the HG is the \textit{token graph} of the underlying lattice~\cite{fabila2012token}. This identification lets us explain several results related to the sign structure of the variational wavefunction and the HAF Hamiltonian. The main contributions of this work are as follows.

\begin{itemize}
\item Any odd cycles in the HG follow from the odd cycles of the physical graph on which the HAF is defined. On a bipartite HG, the two-coloring gives the MSR, while odd cycles in non-bipartite HG obstruct the construction of any sign assignment that satisfies antiferromagnetic alignment on every edge.

\item Building on the correspondence between the Heisenberg Hamiltonian and the token-graph Laplacian~\cite{Osborne2006,apte2025conjectured}, we write $H_\ell=E_F I - L_\ell$ for the \JJ HAF. Here $H_\ell$ is the Hamiltonian restricted to sector $\ell$, $E_F$ is the fully polarized energy, $I$ is the identity matrix and $L_\ell$ is the coupling-weighted HG Laplacian. We
show that computing the GS energy is equivalent to solving a spectrally
relaxed \mct on the HG, in which the signs act as classical Ising spins situated at the vertices of the HG.

\item The $SU(2)$ invariance of the Heisenberg Hamiltonian
yields the following results on the HG Laplacian spectrum: the ladder operators $\hat S^\pm_{\rm tot}$ connect the eigenvectors of Laplacians of neighboring sectors, so the Laplacian spectrum of the HG in sector $\ell-1$ is contained in that of sector $\ell$ for $\ell \le N/2$. 
This provides a physical explanation for a known result on token graph Laplacian spectrum~\cite{ruy_laplacian, apte2025conjectured} and demonstrates that quantum mechanics is a valid way to study the token graph Laplacian. The result also implies that, for an even number of spins, the GS is attained in the zero-magnetization sector.

\item  At fixed amplitudes, the phase-dependent energy is an amplitude-dressed XY model on the HG, and binary phases reduce it to a weighted \mct on HG. For a nondegenerate GS, the Goemans--Williamson relaxation~\cite{goemans1995improved} reproduces the GS signs exactly in time polynomial in the size of the HG, providing a rigorous counterpart to some of the observations used by~\textcite{Westerhout:2022lil} to conclude that the energy landscape is \textit{non-glassy} and \textit{very mildly frustrated}.

\item When amplitudes and phases are parameterized by different sets of variational parameters, VMC acts as an evolving \rt~\cite{burer2002rank} relaxed \mct, whose nonconvex landscape the phase ansatz inherits. This correspondence with the \rt relaxation method also offers a possible explanation for why sign optimization is easier on bipartite lattices, without imposing MSR as a prior.

\item The same odd cycles in non-bipartite HG that prevent antiferromagnetic sign alignment on all edges, also prevent any real diagonal unitary transformation from making the HAF Hamiltonian stoquastic.

\item The average sign, the ratio of the sector partition function to that of its sign-free counterpart, is given by a temperature-dependent Estrada spectral bipartivity~\cite{Estrada} of the HG that retains the Hamiltonian's diagonal part.
\end{itemize}

This HG structure has appeared independently in several contexts. Related constructions include double-vertex graphs~\cite{alavi1991double}, symmetric graph powers~\cite{Audenaert2007Symmetric}, quantum-inspired graph invariants~\cite{rudolph2002constructing}, superlattices~\cite{Nagaoka1966}, and token graphs~\cite{fabila2012token}. Osborne~\cite{Osborne2006} related the XY and Heisenberg Hamiltonians in fixed-magnetization sectors to adjacency and Laplacian operators on graph wedge products, and \textcite{Ouyang2019SpectralBounds} used symmetric products to bound the Heisenberg ferromagnet’s spectrum. None of these studies addressed the sign structure of the many-body wavefunction, which we study here.

The paper is organized as follows. Section~\ref{sec:hilbert_graph} introduces the HG formulation of the HAF, and Section~\ref{sec:hg_topology} studies its structure. Section~\ref{sec:energy_functional} connects the GS energy functional to a relaxation of \mct, while Sec.~\ref{sec:relaxation} discusses optimization methods. Section~\ref{sec: sign learning in vmc} analyzes the non-convexity of the VMC optimization landscape for a particular type of ansatz. In section~\ref{sec:sign_structure_MC}, we show the relation between the stoquastic Hamiltonian and HG \mct, and in section~\ref{sec:average_sign} establish the relation between the average sign in quantum Monte Carlo and the Estrada bipartivity. Section~\ref{sec:conclusion} is for the conclusion and outlook.

\section{Hilbert Graph Formulation of the HAF}\label{sec:hilbert_graph}
In this section, we show that the computational-basis configurations of a HAF in each fixed-magnetization sector carry a graph structure and study its basic structural properties. We then show that, after a constant energy shift, the sector Hamiltonian is the negative weighted
Laplacian of this graph.

\subsection{Hilbert Graph Construction}\label{sec:hilbert_graph_construction}

We represent the physical lattice by a finite, simple, undirected connected graph $G = (V,E)$, where $V$ and $E$ are the vertex and edge sets, respectively, and $N = |V|$ is the number of vertices. Each vertex $i \in V$ carries a spin-$\tfrac12$ degree of freedom $\hat{\mathbf S}_i = \tfrac12 \hat{\boldsymbol\sigma}_i$.

We consider the $J_1$--$J_2$ Heisenberg antiferromagnet (HAF), in which the edge set decomposes as $E=E_1\sqcup E_2$,  where $E_1$ and $E_2$ are fixed by the lattice geometry and contain the nearest-neighbor (NN) and next-nearest-neighbor (NNN) bonds, respectively. We therefore write the graph $G$ in terms of its NN and NNN spanning subgraphs as
\[
G = \bigcup_{r=1}^2 G_r, \quad G_r=(V,E_r).
\]
Here $G_1$ is the NN graph, $G_2$ is the NNN graph, and $G$ is the full interaction graph. Edges in $E_1$ carry coupling $J_1$, while edges in $E_2$ carry $J_2$. For antiferromagnetic couplings $J_1,J_2>0$, the Heisenberg Hamiltonian is
\be\label{eq:Ham}
\hat H = J_1\sum_{\{i,j\}\in E_1}\hat{\mathbf S}_i\cdot \hat{\mathbf S}_j
+ J_2\sum_{\{i,j\}\in E_2}\hat{\mathbf S}_i\cdot \hat{\mathbf S}_j .
\ee
We now separate $\hat H$ into its diagonal Ising part and off-diagonal exchange part, $\hat{H} = \hat{H}_{c} + \hat{H}_{q}$, where
\begin{align}
\hat{H}_{c} &= \sum_{r=1}^{2} J_r\,\hat{H}^{zz}_{r}, \quad
  & \hat{H}^{zz}_{r} &= \sum_{\{i,j\}\in E_r} \hat{S}^z_i \hat{S}^z_j, \label{eq:Hzz-r} \\
\hat{H}_{q} &= \sum_{r=1}^{2} \frac{J_r}{2}\,\hat{H}^{\pm}_{r}, \quad
  & \hat{H}^{\pm}_{r} &= \sum_{\{i,j\}\in E_r} \hat{S}^+_i \hat{S}^-_j + \hat{S}^-_i \hat{S}^+_j. \label{eq:Hpm-r}
\end{align}
Here $\hat H_r^{zz}$ is diagonal in the computational basis, while the off-diagonal term exchanges an antiparallel pair, $(\uparrow_i\downarrow_j\leftrightarrow\downarrow_i\uparrow_j)$,
on a bond in $E_r$. We refer to this elementary exchange as a
Heisenberg flip (HF).

We label each state by a spin configuration \mbox{$\sigma\in\{\downarrow,\uparrow\}^{N}$}. We denote the spin at $i$'th coordinate of $\sigma$ as $\sigma^{(i)}$. If a single HF on a bond $\{i, j\}$ in $E_r$ transforms $\sigma$ into $\tau$, we refer to $\tau$ as a range-$r$ neighbor of $\sigma$. We use the notation $\sigma \overset{ij}{\rightleftharpoons} \tau$ to refer to the fact that exchanging spins of sites $i$ and $j$ changes $\sigma$ to $\tau$ and vice versa. The set of range-$r$ neighbors of $\sigma$ HF is then
\be\label{def: N_r}
\mc N_r(\sigma) = \{\tau: \sigma \overset{ij}{\rightleftharpoons} \tau \text{ for } \{i, j\} \in E_r\}.
\ee
Since a single HF changes two sites, the pair $\{\sigma,\tau\}$
determines the flipped physical bond $\{i,j\}$ uniquely. Because
$E_1\cap E_2=\emptyset$, the induced sets are also disjoint:
$\mc N_1(\sigma) \cap \mc N_2(\sigma) = \emptyset$. We define the set of all range-$r$ neighbors as
\[
\mc N(\sigma) = \bigsqcup_{r=1}^2 \mc N_r(\sigma).
\]
HFs conserve $\hat{S}^{z}_{\mathrm{tot}}=\sum_{i\in V}\hat{S}^{z}_{i}$, or equivalently the number of up spins in a spin configuration $\sigma$, which we denote by $n_\uparrow(\sigma)$. Two basis states can therefore be joined by an HF only if they carry the same value of $n_{\uparrow}(\sigma)$. For HAFs, $n_\uparrow$ is related to the magnetization $m$ by
\be\label{eq:magnetization_sector}
n_\uparrow (\sigma) = m(\sigma) + \frac{N}{2}, \quad m(\sigma) = \braket{\sigma | \hat S^z_\mathrm{tot} | \sigma}.
\ee
Fixing $n_{\uparrow}(\sigma) = \ell$ gives a magnetization sector that is closed under HFs,
\be\label{def: V_l}
\mc V_{\ell} =\left\{ \sigma \in \{\downarrow, \uparrow \}^{N} : n_{\uparrow}(\sigma) = \ell \right\}, \quad \ell = 0,1,\cdots,N.
\ee
The range-$r$ edges inside this sector are
\be\label{def: E_l}
\mc E_{\ell, r} = \{\{\sigma,\tau\}: \sigma \in \mc V_{\ell} \text{ and } \tau \in \mc N_r(\sigma) \}.
\ee
These edge sets are also disjoint, $\mathcal E_{\ell, 1}\cap\mathcal E_{\ell, 2}=\emptyset$. We write the full edge set in sector $\ell$ as
\be\label{eq:hg-edges}
\mc E_\ell = \bigsqcup_{r=1}^2 \mc E_{\ell, r}.
\ee
Therefore, by fixing the number of up spins, we can define a graph through HFs on the vertices. We call this the Hilbert graph (HG) of the $J_1$--$J_2$ HAF in sector $\ell$, and represent it as
\be\label{eq:Gamma_l}
\Gamma_\ell (G) = (\mc V_\ell, \mc E_\ell)
\ee
Because each sector is closed under the HF operation, the full computational-basis HG is the disjoint union of the sector graphs,
\be\label{eq:full_Hilbert_graph}
\Gamma_{\rm full}(G)=\bigsqcup_{\ell=0}^N \Gamma_\ell(G).
\ee
Since different sectors share no vertices, we may analyze each $\Gamma_\ell(G)$ independently. For a physical graph $G$ with $N$ vertices, a vertex of $\Gamma_\ell(G)$
is a choice of $\ell$ sites carrying up spins. Thus $|\mc V_\ell| = \binom{N}{\ell}$ and $\sum_{\ell=0}^N |\mc V_\ell| = 2^N$ is the total number of computational basis states. Hence, for a fixed-$\ell$ sector, the number of HG vertices counts the dimension of the Hilbert space in that sector. 

The fully down-polarized state $\ket{\downarrow,\cdots,\downarrow}$ is the unique vertex of $\Gamma_0(G)$, and we take this as the reference state. A configuration with $\ell$ up spins carries $\ell$ \textit{magnons} above this vacuum. The first nontrivial case is the one-magnon sector, for which
\be\label{eq:one_magnon_hg_tg}
\Gamma_1 (G) \cong G.
\ee
For fixed $\ell$ and large $N$, one has
\[
\binom{N}{\ell}\sim N^{\ell}/\ell!.
\]
Thus the one-magnon HG has $N$ vertices, while the two-magnon HG has
$\mc O(N^2)$ vertices. More generally, if $\ell$ is held fixed as
$N$ grows, the HG grows only polynomially with system size. At finite magnon density, $\alpha=\ell/N$ with $0<\alpha<1$, Stirling’s approximation gives
\[
\binom{N}{\alpha N}\sim
\frac{e^{N h(\alpha)}}{\sqrt{2\pi N\alpha(1-\alpha)}},
\]
where $h(\alpha)=-\alpha\ln\alpha-(1-\alpha)\ln(1-\alpha)$. Thus, HGs at finite magnon density grow exponentially with $N$. The largest sector is the zero-magnetization sector where $\ell = N/2$.

\begin{figure*}[htbp]
\centering
\includegraphics[width=1.0\textwidth]{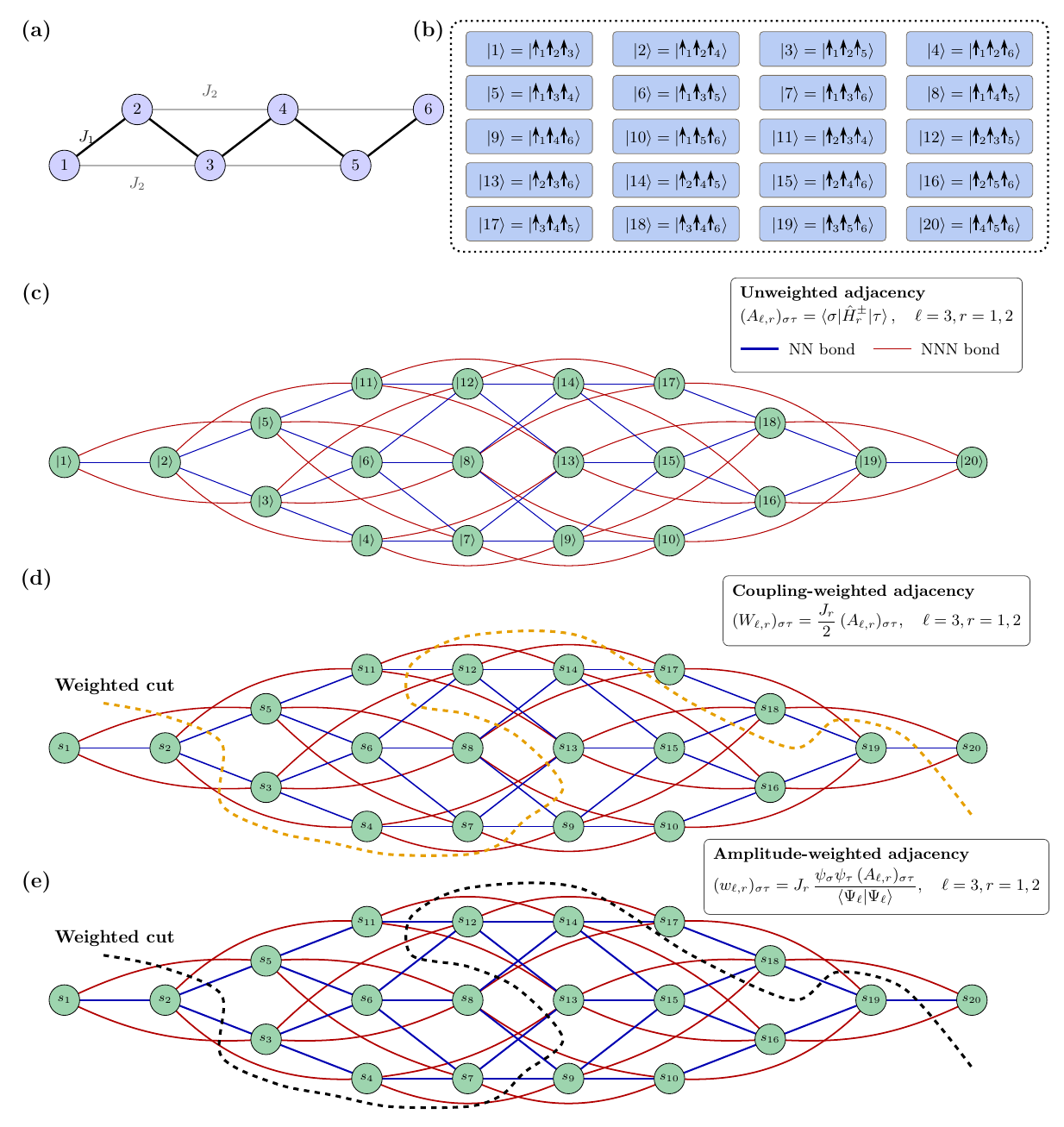}
\caption{Schematic representation of the mapping from a physical lattice to its HG and the associated spectrally relaxed \mct. \textbf{(a)} $J_1$--$J_2$ HAF on a six-site chain with open boundary conditions. \textbf{(b)} The zero-magnetization computational basis, consisting of twenty spin configurations. \textbf{(c)} The corresponding HG is constructed from the off-diagonal matrix elements of the $J_1$--$J_2$ Hamiltonian, with the adjacency matrix in Eq.~\eqref{eq:unweighted_adj2}.  \textbf{(d)} The same HG with each edge carrying its coupling weight $J_r/2$ from Eq.~\eqref{eq:full_weighted_adj_matrix} and each vertex carrying a binary variable $s_\sigma=\pm 1$. A candidate weighted cut is shown as the dashed yellow curve. \textbf{(e)} The GS computation as the spectral relaxation of the \mct in (d), in which each $s_\sigma$ is replaced by $\psi_\sigma s_\sigma$. With the GS amplitudes $\psi_\sigma$ held fixed, each edge carries the amplitude-dressed weight of Eq.~\eqref{def: amplitude weighted adjacency} and each vertex carries a binary phase $s_\sigma=e^{i\phi_\sigma}$, $\phi_\sigma\in\{0,\pi\}$. Minimizing the energy is then equivalent to maximizing the total dressed weight of the cut edges; the black dashed curve shows a candidate cut. Fixing the amplitudes to their GS values thus turns the sign optimization into the phase reconstruction problem (\prp).
}
\label{fig:HG}
\end{figure*}

Intermediate scalings of $\ell$ give intermediate growth. For example,
when $\ell\sim\sqrt{N}$, the HG size grows faster than any power of $N$
but slower than $\exp(cN)$ for fixed $c>0$. Thus, the scaling depends on
the sector: fixed-$\ell$ sectors grow polynomially, whereas sectors with
$\ell=\alpha N$ grow exponentially. The exponential growth is largest
near the center of the magnetization tower. Moreover, a global spin flip maps the $\ell$ sector to the $(N-\ell)$ sector, giving the duality
\be\label{eq:duality}
\Gamma_{\ell}(G) \cong \Gamma_{N-\ell}(G).
\ee
For even $N$, the \textit{self-dual} sector is $\ell=N/2$, which coincides with the zero-magnetization sector. Fig.~\ref{fig:HG} illustrates the construction of $\Gamma_{N/2}(G)$ for a six site chain $J_1$--$J_2$ HAF.

The HG is built from the nonzero off-diagonal matrix elements of the Hamiltonian $\hat H$ in Eq.~\eqref{eq:Ham}, so both its vertex set and its connectivity depend on the basis in which it is written. We have worked throughout in the computational basis, which lets us identify it as a token graph~\cite{fabila2012token}.

In a fixed magnetization sector $\ell$, represent each up spin by a \textit{token} and each down-spin site by an \textit{unoccupied} vertex of $G$. A configuration $\sigma\in\mc V_\ell$ is then a placement of $\ell$ indistinguishable tokens on $V$. An HF acts only on an antiparallel pair, so along a bond $\{i,j\} \in E$ it moves a single token onto an adjacent unoccupied vertex and leaves every other token where it is. This is the adjacency rule of
the $\ell$-token graph $F_\ell(G)$~\cite{fabila2012token}. Hence,
\be\label{eq:HG-token-graph}
\Gamma_\ell(G)\cong F_\ell(G),
\ee
where $F_\ell(G)$ denotes the $\ell$-token graph of $G$. Restricting the
HFs to bonds in $E_1$ or $E_2$ gives
\[
\Gamma_\ell(G_1)\cong F_\ell(G_1), \qquad \Gamma_\ell(G_2)\cong F_\ell(G_2).
\]
The same correspondence therefore holds for the NN and NNN interaction
graphs separately. The relation $\Gamma_1(G)\cong G$ is the
$\ell=1$ case, since $F_1(G)\cong G$. 

Under a general unitary transformation, the matrix elements of $\hat H$ may be mixed, and the resulting HG need not have the connectivity of $F_\ell(G)$. The token graph therefore serves as the \textit{bare} HG, the structure the HAF carries before any basis transformation.  Sec.~\ref{sec:sign_structure_MC} discusses basis transformations, where the diagonal unitary transformations preserve the HG connectivity and hopping magnitudes, changing only the signs of the corresponding off-diagonal matrix elements. Identifying the HG as the token graph lets us express the Hamiltonian in terms of the graph Laplacian, which we discuss next.

\subsection{Hamiltonian as the Hilbert Graph Laplacian}\label{subsec:Ham-Lap}

At the operator level, the exchange term in Eq.~\eqref{eq:Hpm-r} acts as the adjacency operator on the HG. For each range $r$, $\hat H_r^{\pm}$ connects exactly those configurations joined by an edge in $\mc E_{\ell, r}$. Its matrix is therefore the range-$r$ adjacency matrix written in the computational basis indexed by $\mc V_\ell$,
\be\label{eq:A_rl}
\left(A_{\ell, r}\right)_{\sigma\tau} = \braket{\sigma | \hat H_r^\pm | \tau} =
\begin{cases}
    1, \quad \{\sigma, \tau\} \in \mc E_{\ell, r} \\
    0, \quad \text{otherwise}.
\end{cases}
\ee
The full unweighted adjacency matrix is
\be\label{eq:unweighted_adj2}
A_\ell = \sum_{r=1}^2 A_{\ell, r}
\ee
Assign weight $J_r/2$ to each range-$r$ HG edge. The resulting coupling-weighted adjacency matrix is
\be\label{eq:full_weighted_adj_matrix}
W_\ell = \sum_{r=1}^2 W_{\ell, r} \, , \quad W_{\ell, r} = \frac{J_r}{2} \, A_{\ell, r}
\ee
The degree matrix of a graph gives us the number of connections each vertex has. For the unweighted adjacency matrix of the HG, the degree matrix is determined by the number of antiparallel bonds. For a configuration $\sigma\in\mathcal V_\ell$,
let
\be\label{eq:anti-parallel-config}
n_r^{\mathrm{ap}}(\sigma) = \left|\left\{\{i,j\}\in E_r:\sigma^{(i)}\neq\sigma^{(j)}\right\}\right|
\ee
be the number of antiparallel range-$r$ bonds. Each such bond supports
one range-$r$ HF, so $n_r^{\mathrm{ap}}(\sigma)$ is the unweighted
range-$r$ degree of $\sigma$. The coupling-weighted degree matrix is therefore
\be\label{eq:degree_matrix}
(D_\ell)_{\sigma\sigma}
=\frac{1}{2}\sum_{r=1}^{2}J_r n_r^{\rm ap}(\sigma)
=\sum_{\tau \in \mc V_\ell} (W_\ell)_{\sigma\tau}.
\ee
Since $\hat S_i^z\hat S_j^z$ has eigenvalue $1/4$ for parallel spins and
$-1/4$ for antiparallel spins, each antiparallel range-$r$ bond
lowers the Ising energy by $J_r/2$ below the ferromagnetic
reference energy, and
\be\label{eq:ising_energy_degree}
\braket{\sigma|\hat H_c|\sigma}=E_F-(D_\ell)_{\sigma\sigma}, \qquad E_F=\frac{1}{4}\sum_{r=1}^{2}J_r|E_r|.
\ee
In the computational basis indexed by $\mc V_\ell$,
Eq.~\eqref{eq:ising_energy_degree} and the definition of
$W_\ell$ give us the diagonal (classical) and the off-diagonal (quantum) parts of the Hamiltonian in sector $\ell$.
\be\label{eq:Hc_equals_degree}
H_{\ell,c}=E_FI-D_\ell,
\qquad
H_{\ell,q}=W_\ell .
\ee
Thus, up to the constant shift $E_F$, the classical part gives
$-D_\ell$, while the quantum part gives the weighted adjacency matrix
$W_\ell$. The matrix representation of the Hamiltonian can then be written as
\be\label{eq:H_degree_adjacency}
H_\ell = E_F I - D_\ell + W_\ell.
\ee
Defining the coupling-weighted Laplacian of the HG as
\be\label{eq:weighted_HG_laplacian}
L_\ell = D_\ell - W_\ell,
\ee
and combining Eqs.~\eqref{eq:H_degree_adjacency} with~\eqref{eq:weighted_HG_laplacian} gives
\be\label{eq:H_laplacian_shifted}
H_\ell=E_FI-L_\ell .
\ee
After shifting the energy origin by the constant $E_F$, we write
\be\label{eq:H_equals_minus_L}
H_\ell\equiv -L_\ell .
\ee
The identity~\eqref{eq:H_equals_minus_L} connects the HAF
Hamiltonian to the Laplacian of the HG. It therefore turns the GS problem
into a spectral problem on the HG: in each sector, the GS energy of the
shifted Hamiltonian is the negative of the largest Laplacian eigenvalue,
and the GS is the corresponding Laplacian eigenvector. We order its eigenvalues as
\[
0=\lambda_0 \leq \lambda_1 \leq \cdots \leq \lambda_{|\mc V_\ell| - 1} .
\]
Eq.~\eqref{eq:H_equals_minus_L} shows that $H_\ell$ and $L_\ell$
have the same eigenvectors, while their eigenvalues have opposite signs.
If the Hamiltonian energies in sector $\ell$ are ordered as
$E_{0}\leq E_{1}\leq\cdots\leq
E_{|\mc V_\ell|-1}$, then
\be\label{eq:spectrum_reversal}
E_{j} = - \lambda_{|\mc V_\ell|-1-j}, \qquad j=0,1,\ldots,|\mc V_\ell|-1.
\ee
The highest Hamiltonian energy is
$E_{|\mathcal V_\ell|-1}=-\lambda_0=0$. The corresponding
Laplacian zero mode is the uniform superposition of all configurations
in $\mathcal V_\ell$. In sector $\ell=0$, there is only one
configuration, the state with all spins down, so this state itself is the zero mode.

For an HG with more than one vertex, the zero eigenvalue is nondegenerate
because the HG is connected~\cite{fabila2012token}, and the second-smallest Laplacian eigenvalue $\lambda_1>0$ is the Fiedler eigenvalue, also called the algebraic connectivity, which measures how well connected the HG is. Since the two spectra are reversed, it fixes the second-highest Hamiltonian energy in sector $\ell$,
\be\label{eq:Fiedler_H_energy}
E_{|\mc V_\ell|-2} = -\lambda_1,
\ee
which is negative of the Fiedler value. The GS energy in sector $\ell$ is the negative of the largest
Laplacian eigenvalue,
\be\label{eq:GS_energy_lmax}
E_{0} = -\lambda_{|\mc V_\ell| - 1}.
\ee

The relation between the Heisenberg Hamiltonian and graph Laplacians in Eq.~\eqref{eq:H_equals_minus_L} appeared in different contexts~\cite{Osborne2006,Ouyang2019SpectralBounds,apte2025conjectured}.  Theorem~\ref{thm:HG--TG-Correspondence} below states the correspondence for coupling-weighted HGs. For the \JJ HAF, $L_\ell$ separates into its NN and NNN contributions.

\begin{theorem}[Hilbert--Token Graph Correspondence]
\label{thm:HG--TG-Correspondence}
Let $G=(V,E)$ be a finite, simple, undirected graph whose edge set is
partitioned into disjoint bond sets,
\[
E = \bigsqcup_r E_r ,
\]
as determined by the lattice geometry. For each $E_r$, let $J_r>0$
denote the corresponding antiferromagnetic coupling. The HAF
Hamiltonian on $G$ is
\[
\hat H = \sum_r J_r \sum_{\{i, j\} \in E_r} \hat {\mathbf S}_i \cdot \hat{\mathbf S}_j.
\]
Fix a sector $\ell\in\{0,1,\ldots,N\}$. Then:
\begin{enumerate}
\item[(i)] As an unweighted graph, the HG in sector $\ell$ is
isomorphic to the $\ell$ token graph of $G$,
\be\label{eq:thm_iso}
\Gamma_\ell(G) \cong F_\ell(G) .
\ee
Here $\Gamma_\ell(G)$ is the HG defined in Eq.~\eqref{eq:Gamma_l}.

\item[(ii)] In the computational basis indexed by $\mathcal V_\ell$,
the Hamiltonian matrix $H_\ell$, after subtracting the ferromagnetic
reference energy, satisfies
\be\label{eq:theorem_H_minus_L}
H_\ell=-L_\ell.
\ee
Here $L_\ell = D_\ell - W_\ell$ is the coupling-weighted HG Laplacian of Eq.~\eqref{eq:weighted_HG_laplacian}, which decomposes as
\be\label{eq:thm_L_decomposition}
L_\ell=\sum_r\frac{J_r}{2} \, L_{\ell, r},
\qquad
L_{\ell, r}=D_{\ell, r}-A_{\ell, r},
\ee
where $D_{\ell,r}$ and $A_{\ell,r}$ are the unweighted range-$r$ degree and adjacency matrices.

\item[(iii)] After the same energy shift, the spectrum of $H_\ell$ is
the negative of the Laplacian spectrum,
\be\label{eq:theorem_spectrum}
\operatorname{spec}(H_\ell)=-\operatorname{spec}(L_\ell).
\ee
In particular, the GS energy in sector $\ell$ is the negative of the largest eigenvalue of the graph Laplacian.
\be\label{eq:theorem_GS_energy}
E_{\min}(H_\ell)=-\lambda_{\max}(L_\ell).
\ee
\end{enumerate}
\end{theorem}

\begin{proof}
Part~(i) follows directly from the TG identification in
Eq.~\eqref{eq:HG-token-graph}.

For part~(ii), Eqs.~\eqref{eq:Hpm-r}
and~\eqref{eq:full_weighted_adj_matrix} show that the exchange part of
the Hamiltonian gives the weighted HG adjacency operator.
Eq.~\eqref{eq:Hc_equals_degree} gives the diagonal contribution as the HG degree operator, together with the ferromagnetic reference energy. Using the definition of the weighted HG Laplacian in Eq.~\eqref{eq:weighted_HG_laplacian}, and subtracting the reference
energy, gives $H_\ell=-L_\ell$.

The decomposition in Eq.~\eqref{eq:thm_L_decomposition} follows by
collecting the contributions from each bond set $E_r$. This proves
part~(ii).

Part~(iii) follows from part~(ii). Multiplication by $-1$ preserves the
eigenvectors and reverses the signs and ordering of the eigenvalues.
Therefore,
\[
\operatorname{spec}(H_\ell)=-\operatorname{spec}(L_\ell).
\]
The smallest eigenvalue of $H_\ell$ corresponds to the largest
eigenvalue of $L_\ell$, giving Eq.~\eqref{eq:theorem_GS_energy}.
\end{proof}
Theorem~\ref{thm:HG--TG-Correspondence} underlies our later study of the sign structure of the variational wavefunction and of the stoquasticity of the HAF Hamiltonian. For now we will use it with the $SU(2)$ symmetry of the Heisenberg Hamiltonian, to show how a property of token graph Laplacian spectrum is connected to the GS for even $N$ living in the zero-magnetization sector.

\subsection{Laplacian Spectrum and GS Magnetization}\label{subsec:laplacian_spectrum}

We have shown that the largest eigenvalue of the graph Laplacian corresponds to the GS energy. In this section, we determine the magnetization sector of the GS. Lieb and Mattis~\cite{LM1} showed that, for an even number of sites, the HAF GS on a bipartite lattice lies in the zero-magnetization sector. Here we show that this also holds for non-bipartite geometries by combining angular-momentum algebra with the spectral inclusion result for token-graph Laplacians~\cite{ruy_laplacian}.

We consider an isotropic Heisenberg model on the physical graph $G=(V,E)$, with bond-dependent couplings $J_{ij}$.
\be\label{def: general Heisenberg Model}
\hat{\bar{H}} = \sum_{\{i, j\} \in E} J_{ij} \, \mathbf{\hat S}_i \cdot \mathbf{\hat S}_j, \quad J_{ij} \in \mathbb{R}
\ee
Using the notation of HF used in Eq.~\eqref{def: N_r}, this Hamiltonian generates the weighted adjacency matrix,
\[
(\bar W_{\ell})_{\sigma\tau} =
\begin{cases}
J_{ij}/2, & \sigma \overset{ij}{\rightleftharpoons} \tau \text{ and } \{i,j\} \in E,\\
0, & \text{otherwise,}
\end{cases}
\]
for sector $\ell$ spin configurations $\sigma,\tau \in \mc V_\ell$. Now, we can think of the couplings $J_{ij}$ as weights on the physical graph $G$. Then the token graph inherits these weights through HF. In the language of Ref.~\cite{fabila2012token}, if the symmetric difference between two token graph vertices connected by an edge is $\{i, j\}$, then that edge inherits the parent graph weight $J_{ij}$ (scaled by a factor of $1/2$). The Laplacian of this weighted token graph, $\bar L_\ell$, is connected to the matrix representation of the Hamiltonian in sector $\ell$, $\bar H_\ell$ as
\be\label{eq: H in terms of L general}
\bar H_\ell = \bar E_F \, I - \bar L_\ell, \qquad \bar{E}_F = \sum_{\{i, j\}\in E} \frac{J_{ij}}{4}
\ee
The Heisenberg Hamiltonian is $SU(2)$ symmetric. Thus, we can find a simultaneous orthonormal eigenbasis of $\hat{\bar{H}}$, $\mathbf{S}^2_\mathrm{tot}$, and $S^z_\mathrm{tot}$, as they are mutually commuting operators. Then, the energy eigenstates are also angular momentum eigenstates labeled as $\ket{E, s, m_s}$, where $E$, $s(s+1)$, and $m_s$ are the eigenvalues of $\hat{\bar{H}}$, $\mathbf{S}^2_\mathrm{tot}$, and $S^z_\mathrm{tot}$, respectively. Each eigenstate belongs to the $\ell = m_s + N/2$ sector~\eqref{eq:magnetization_sector} and can be expressed as a superposition of basis states in $\mc V_\ell$, making it an eigenvector of $\bar H_\ell$. The ladder operators $S^\pm_\mathrm{tot}$ commute with the Hamiltonian; hence, applying them to the eigenstates generates states with the same energy and spin but different magnetization. In other words, the ladder operators move the eigenstates from one sector to another
\[
S^\pm_\mathrm{tot} \ket{E, s, m_s} \equiv \ket{E, s, m_s \pm 1}, \qquad |m_s\pm 1| \le s.
\]
All eigenstates in a single multiplet have the same energy; therefore, an eigenstate with spin $s$ has degeneracy at least $2s + 1$.
\begin{theorem}[Nested Spectrum]
\label{thm: nested spectrum}
The spectrum of $\bar H_\ell$ is nested such that the sector closer to zero magnetization contains the spectra of sectors farther away.
\begin{align*}
&\mathrm{spec}(\bar H_\ell) \subseteq \mathrm{spec}(\bar H_{\ell+1}), &1\leq\ell< N/2, \\
&\mathrm{spec}(\bar H_\ell) \supseteq \mathrm{spec}(\bar H_{\ell+1}), &N/2\leq\ell\leq N.
\end{align*}
\end{theorem}
\begin{proof}
Assume we are in sector $\ell$. Since $m_s = \ell - N/2$ every eigenvector, $\ket{E, s, m_s}$, of $\bar H_\ell$ has spin $s \geq |\ell - N/2|$. For $1 \leq \ell < N/2$, we can use $S^+_\mathrm{tot}$ to move all the eigenvectors to the sector $\ell + 1$. These eigenvectors have the same energy eigenvalues in sector $\ell + 1$ as in sector $\ell$, since the ladder operators commute with the Hamiltonian. For $\ell \geq N/2$ the same arguments using $S^-_\mathrm{tot}$ lead to $ \mathrm{spec}(\bar H_{\ell+1}) \subseteq \mathrm{spec}(\bar H_{\ell})$.
\end{proof}
\begin{corollary}[GS Magnetization]
The GS of $\bar H_{\lfloor N / 2 \rfloor}$ corresponds to the GS of $\hat{\bar H}$, in other words, the $\ell = \lfloor N/2 \rfloor$ sector contains a GS of $\hat{\bar H}$. For an even number of spins, this is the zero magnetization sector.
\end{corollary}
\begin{proof}
The corollary follows directly from Theorem~\ref{thm: nested spectrum} since $\mathrm{spec}(\bar H_{\lfloor N/2 \rfloor})$ contains the spectra of all other sectors. No matter the magnetization of GS, we can always bring it to $\lfloor N/2 \rfloor$ sector by repeated application of a ladder operator without changing energy eigenvalue. The choice of $\ell = \lfloor N/2 \rfloor$ is not unique, the $\ell = \lceil N/2 \rceil$ sector works just as well for odd number of spins due to duality~\eqref{eq:duality}.
\end{proof}

Since the Hamiltonian and the Laplacian are related by Eq.~\eqref{eq: H in terms of L general}, $\mathrm{spec}(\bar L_\ell) \subseteq \mathrm{spec}(\bar L_{\ell+1})$ for $1\leq\ell<N/2$. This theorem was proven in Ref.~\cite{ruy_laplacian} for the unweighted case, i.e., $J_{ij}=2$, and was extended to the weighted case in Ref.~\cite{apte2025conjectured}. However, neither work used the $SU(2)$ invariance to prove this result. Our approach shows a physical reason for why the nested spectrum property is true, that it is a consequence of the $SU(2)$ symmetry and angular momentum eigenstates, and thus provides an example of how purely quantum mechanical approach can be used to study the token graph Laplacian.

Theorem~\ref{thm: nested spectrum} makes no claim about whether there exists a singlet GS or if all GS are singlet, for even number of spins. All GS being singlet is equivalent to the following strict inequality in token graph language.
\[
\lambda_\mathrm{max}(\bar L_{N/2-1}) < \lambda_\mathrm{max}(\bar L_{N/2})
\]
According to \textcite{LM1}, this is true for connected bipartite graphs with antiferromagnetic couplings, when the bipartition divides the graph into two parts with equal number of vertices. However, no analytical proof exists for commonly used non-bipartite graphs, such as the periodic square lattice with NN and NNN edges for $J_2/J_1 > 0.5$ (The $J_2/J_1 \leq 0.5$ case was proven by \textcite{Tanaka}). We believe insights from token graph literature can help address this analytically.

\section{Structure of the Hilbert graph}\label{sec:hg_topology}

We now use the definition of the HG and its adjacency matrices in Eqs.~\eqref{eq:A_rl} and~\eqref{eq:unweighted_adj2} to study HG properties relevant to our work: bipartiteness, triangle count, and the Estrada bipartivity index~\cite{Estrada}. The first provides a structural condition, while the latter two characterize odd cycles in HG.

\subsection{Bipartiteness Inheritance}
The HG inherits both bipartiteness and odd cycles from the parent interaction graph $G$; a bipartition of $G$ yields a two-coloring of $\Gamma_\ell(G)$, and an odd cycle in $G$ yields an odd cycle in the HG. The following theorems prove these two inheritance properties.

\begin{theorem}[Cycle inheritance]\label{thm: cycle inheritance}
Let $G=(V,E)$ be a finite simple undirected graph and 
$1\leq\ell\leq |V|-1$. Then, any cycle in $G$ can be mapped to a cycle of the same length in
$\Gamma_\ell(G)$.
\end{theorem}

\begin{proof}
Consider a cycle $i_1, i_2, \ldots, i_L$ in $G$, where $i_k$ are unique vertices of $G$. First we assume $L \leq \ell + 1$ and place a down spin at $i_1$ and up spin on the rest of the cycle. The remaining vertices are always assumed to be filled in a way that satisfies the magnetization sector. Using HF we can transfer the down spin at $i_1$ as $i_1 \to i_2 \to \ldots \to i_L \to i_1$, which creates a cycle in $\Gamma_\ell(G)$ of length $L$. Now we consider $L > \ell + 1$ and fill $i_1, i_2, \ldots, i_\ell$ with up spins and $i_{\ell + 1}, \ldots, i_L$ with down spins. We move the up spin at $i_\ell$ as $i_\ell \to i_{\ell + 1} \to \ldots \to i_L$, taking $L - \ell$ HFs. We then move the up spins at $i_1, i_2, \ldots, i_{\ell-1}$ as $i_k \to i_{k+1}$ starting from $i_{\ell-1}$ and stopping at $i_1$. This takes another $\ell - 1$ HFs. Now we have a down spin at $i_1$ and an up spin at $i_L$, which we exchange using another HF. This creates a cycle of length $(L - \ell) + (\ell - 1) + 1 = L$ in $\Gamma_\ell(G)$.
\end{proof}
The theorem above shows that any odd cycles in $G$ yield odd cycles in $\Gamma_\ell(G)$. It follows that if $G$ is non-bipartite, $\Gamma_\ell(G)$ must also be non-bipartite. Nevertheless, this construction does not account for all the odd cycles in the HG; in general, the HG may have additional odd cycles which are not direct images of cycles of the same length in $G$. The next theorem, which was first proved in \cite{fabila2012token}, states that if $G$ contains no odd cycles, then $\Gamma_\ell(G)$ cannot have any odd cycles.

\begin{theorem}[Bipartiteness inheritance]
\label{thm:bipartite_inheritance}
Let $G=(V,E)$ be a finite simple undirected graph.  For
$1\le \ell\le |V|-1$, $\Gamma_\ell(G)$ is bipartite if and only if
$G$ is bipartite.
\end{theorem}

\begin{proof}
Theorem~\ref{thm: cycle inheritance} shows that if $G$ is non-bipartite then so is $\Gamma_\ell(G)$. Now we only need to show that if $G$ is bipartite then so is $\Gamma_\ell(G)$. Suppose $G$ is bipartite. Then its vertex set is a union of two non-empty disjoint sets, or sublattices, $V=V_A \sqcup V_B$, such that every edge of $G$ has one endpoint in $V_A$ and the other in $V_B$. For a configuration $\sigma\in\mathcal V_\ell$ define the number of up spins in $V_A$ as
\be\label{eq:N_A}
N_A^\uparrow(\sigma) := \lvert \{i\in V_A:\sigma^{(i)}=\,\uparrow\} \rvert,
\ee
and its parity
\be\label{eq:msr_sign_one}
\Pi(\sigma):= (-1)^{N_A^\uparrow(\sigma)}.
\ee
Let $\{\sigma,\tau\}\in\mc E_{\ell}$ be an HG edge.  Then $\tau$ is
obtained from $\sigma$ by an HF on some bond $\{i,j\}\in E$, with $i\in V_A$ and $j\in V_B$. For an HF, the two spins on this bond are antiparallel, meaning only one spin is up. After the flip, the up spin moves from either $V_A$ to $V_B$ or $V_B$ to $V_A$.  Hence
$N_A^\uparrow$ changes by $\pm 1$, and therefore
\[
\Pi(\tau)= - \Pi(\sigma)
\]
on every HG edge. Thus, $\Pi$ gives a two-coloring of $\Gamma_\ell(G)$, making it bipartite.
\end{proof}

The parity $\Pi(\sigma)= (-1)^{N_A^\uparrow(\sigma)}$ is the well-known MSR, which follows from Theorem~\ref{thm:bipartite_inheritance}. Thus, when the physical graph $G$ is bipartite, the MSR assigns signs to the HG vertices that define its bipartition. Same-sublattice couplings, such as $J_2$ bonds on the square lattice, connect HG vertices with the same $\Pi(\sigma)$ value, so this coloring is no longer proper. 

The non-bipartiteness therefore manifests in the HG through triangles and higher-order odd cycles. We next count triangles in the HG.

\subsection{Triangle Count}
\label{sec:triangles}

The shortest odd cycles in the HG are triangles, which we read from the unweighted adjacency matrix $A_\ell$ in Eq.~\eqref{eq:unweighted_adj2}. Since $\left(A_\ell^{3}\right)_{\sigma\sigma}$ counts closed walks of length three based at $\sigma$, each triangle contributes six closed walks of length three: three starting vertices and two traversal directions. The number of triangles in the HG is therefore~\cite{west2001graphtheory}
\be\label{eq:triangle}
N_{\triangle} (\Gamma_\ell) = \frac{1}{6}\,\operatorname{Tr}\!\left(A_{\ell}^{3}\right). 
\ee
A positive value of $N_{\triangle} (\Gamma_\ell)$ certifies that the HG contains triangles and is therefore not bipartite; by Theorem~\ref{thm:bipartite_inheritance}, this can occur only when the physical lattice $G$ is itself non-bipartite. 

\begin{table*}[t]
\centering
\small
\renewcommand{\arraystretch}{1.2}
\begin{tabular*}{\textwidth}{@{\extracolsep{\fill}} l c c c ccc ccc @{}}
\toprule
& & & & \multicolumn{3}{c}{PBC} & \multicolumn{3}{c}{OBC} \\
\cmidrule(lr){5-7} \cmidrule(lr){8-10}
Lattice & Size & $|\mc V|$ & $J_2/J_1$ & $\beta$ & $N_{\triangle}$ & $|\mc E|$ & $\beta$ & $N_{\triangle}$ & $|\mc E|$ \\
\midrule
\multirow{2}{*}{Chain} & \multirow{2}{*}{$16$} & \multirow{2}{*}{$12{,}870$}
  & zero     & $1.00$ & $0$         & $54{,}912$  & $1.00$ & $0$         & $51{,}480$  \\
& & & non-zero & $0.51$ & $54{,}912$  & $109{,}824$ & $0.51$ & $48{,}048$  & $99{,}528$  \\
\midrule
\multirow{2}{*}{Square} & \multirow{2}{*}{$4\times4$} & \multirow{2}{*}{$12{,}870$}
  & zero     & $1.00$ & $0$         & $109{,}824$ & $1.00$ & $0$         & $82{,}368$  \\
& & & non-zero & $0.50$ & $219{,}648$ & $219{,}648$ & $0.50$ & $123{,}552$ & $144{,}144$ \\
\midrule
\multirow{2}{*}{Triangular} & \multirow{2}{*}{$4\times4$} & \multirow{2}{*}{$12{,}870$}
  & zero     & $0.50$ & $109{,}824$ & $164{,}736$ & $0.50$ & $61{,}776$ & $113{,}256$ \\
& & & non-zero & $0.50$ & $878{,}592$  & $329{,}472$ & $0.50$ & $233{,}376$ & $185{,}328$ \\
\bottomrule
\end{tabular*}
\caption{Hilbert graph diagnostics under periodic (PBC) and open (OBC)
boundary conditions for the $J_1$--$J_2$ HAF on three lattices: the
$16$-site chain, the $4\times4$ square, and the $4\times4$ triangular
lattice. $N_{\triangle}(\Gamma_{N/2})$ is the number of triangles in the HG,
Eq.~\eqref{eq:triangle}, and $\beta$ is the Estrada bipartivity index,
Eq.~\eqref{eq:bipartivity}.}
\label{tab:hg_triangle_count}
\end{table*}

In Table~\ref{tab:hg_triangle_count}, we show $N_{\triangle}$, computed
from $A_\ell$ using Eq.~\eqref{eq:triangle}, for the \JJ chain, square, and triangular-lattice HGs under periodic and open boundary conditions in the zero-magnetization sector at $N=16$. 

We note that in the chain and square lattice cases, the HG contains no triangles when $J_{2}=0$. This is consistent with Theorem~\ref{thm:bipartite_inheritance}, showing that if the parent graph is bipartite, its HG is as well. In the triangular lattice, however, the HG contains triangles when $J_2 = 0$. This shows that the HG inherits triangles from its parent graph. For non-zero $J_2$, all three HGs inherit triangles from their respective parent graph. 

We later show that the signs of the variational wavefunction, or equivalently the signs of the off-diagonal part of the HAF Hamiltonian, define an antiferromagnetic Ising model on the HG.  For non-bipartite geometries, these triangles are the origin of geometric frustration in the HG, which prevents a Marshall-type phase prior for the wavefunction. Triangles on the HG can be sufficient to create geometric frustration, but not necessary. Following Theorem~\ref{thm: cycle inheritance}, an HG may contain higher-order odd cycles. 

The triangle count provides a one-sided witness: $N_{\triangle}>0$ proves that the HG is non-bipartite, whereas $N_\triangle=0$ does not prove the bipartiteness. In the next section, we show that the Estrada bipartivity index accounts for these higher-order odd cycles.

\subsection{Estrada Bipartivity Index}\label{subsec:estrada}
The Estrada bipartivity index~\cite{Estrada} is a spectral measure based on the weighted proportion of even closed walks to all closed walks in a graph. It equals one for a bipartite graph, where every closed walk has even length, and falls below one when odd closed walks are present.

The number of walks from $\sigma$ to $\tau$ of length $k$ in $\Gamma_\ell$ is equal to $(A_\ell^k)_{\sigma\tau}$. Therefore, $\operatorname{Tr}(A^k_\ell) = \sum_{\sigma \in \mc V_\ell} (A^k_\ell)_{\sigma\sigma}$ gives us the total number of closed walks of length $k$. However, this number can diverge as $k$ gets larger. \textcite{Estrada} used a factor of $1/k!$ which accounts for this divergence. Then, a measure of all even closed walks is
\[
\operatorname{Tr} \cosh A_\ell = \sum_{k=0}^\infty \frac{1}{(2k)!} \operatorname{Tr}(A^{2k}_\ell)
\]
and a measure of all odd closed walks is,
\[
\operatorname{Tr} \sinh A_\ell = \sum_{k=0}^\infty \frac{1}{(2k + 1)!} \operatorname{Tr}(A^{2k + 1}_\ell).
\]
Since $\operatorname{Tr}(A_\ell) = 0$ by definition, the first, or leading, term in the expansion of $\operatorname{Tr} \sinh A_\ell$ is $N_\triangle$~\eqref{eq:triangle}. The contributions from higher-order odd closed walks are also included, but with smaller weight factors. Now, the Estrada bipartivity is defined as the proportion of even to total closed walks.
\[
\beta(\Gamma_\ell) = \frac{\operatorname{Tr}\cosh A_\ell}{\operatorname{Tr}\cosh A_\ell + \operatorname{Tr}\sinh A_\ell} = \frac{\operatorname{Tr} \cosh A_\ell}{\operatorname{Tr} \, e^{A_\ell}}
\]
In terms of eigenvalues of $A_\ell$, denoted by $\{\lambda_i\}$ we can express the bipartivity as,
\be\label{eq:bipartivity}
\beta(\Gamma_\ell) = \frac{\sum_i \cosh \lambda_i}{\sum_i e^{\lambda_i}}
\ee
Unlike the triangle count, $\beta$ includes closed walks of every
length. It satisfies~\cite{Estrada}
\begin{equation}
\label{eq:beta_range}
\tfrac{1}{2} < \beta(\Gamma_\ell) \le 1 .
\end{equation}
The upper bound $\beta = 1$ is attained when all the closed walks are even, and the contribution from odd closed walks is zero, i.e., $\operatorname{Tr} \sinh A_\ell = 0$. This can only happen when $\operatorname{Tr}(A_\ell^{2k+1}) = 0 \  \forall k$, meaning there are no odd closed walks in the graph. Therefore, $\beta(\Gamma_\ell(G)) = 1$ implies $\Gamma_\ell(G)$ is bipartite and by Theorem~\ref{thm:bipartite_inheritance} it also implies that $G$ is bipartite. Conversely, any value below unity signals at least one odd closed walk, so no global two-coloring of $\Gamma_{\ell}(G)$ exists.

Table~\ref{tab:hg_triangle_count} shows $\beta$ for the
zero-magnetization HGs of the \JJ HAF on the chain, square,
and triangular lattices. At $J_2=0$, the chain and square HGs are
bipartite and have $\beta=1$, whereas the triangular lattice HG is
non-bipartite and has $\beta\simeq0.50$. For nonzero $J_2$, $\beta$
drops to $0.51$ for the chain and to approximately $0.50$ for the
square, while remaining near $0.50$ for the triangular lattice. To
the precision shown, these values are unchanged between periodic and
open boundary conditions.

In the next section, we write the HAF energy directly as a functional on the HG in a fixed-magnetization sector. This form shows how the HG topology constrains phase reconstruction of a variational wavefunction and connects it to \mct.

\begin{figure*}[htbp]
    \centering
    \includegraphics[width=1.0\textwidth]{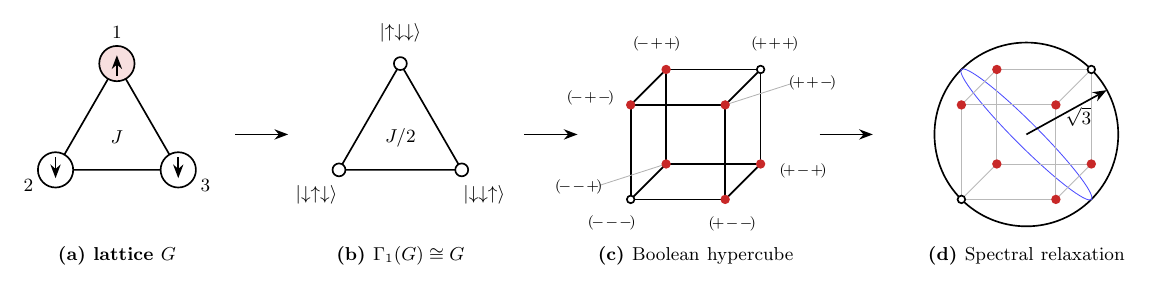}
\caption{Spectral relaxation of the discrete \mct problem on the
one-magnon HG of a triangle. \textbf{(a)} A single up spin on the three-site antiferromagnetic lattice. \textbf{(b)} In the one-magnon sector, the HG is a triangle, $\Gamma_1(G)\cong G$, with edge weight $J/2$.
\textbf{(c)} The binary signs $x_\sigma=\pm1$ form the eight vertices of the Boolean cube. The six red vertices give the maximum cuts of the
triangle, while the two white vertices do not cut any edge.
\textbf{(d)} Spectral relaxation replaces these eight discrete vectors by the sphere $\lVert\mathbf{x}\rVert^2=3$. The cube is retained in gray to show that the original sign assignments remain inside the relaxed
space. The blue circle marks where the maximum is attained,
$\sum_{\sigma} x_{\sigma} = 0$, which no cube corner reaches.}
\label{fig:spectral_relaxation}
\end{figure*}

\section{The Energy Functional and \mct}\label{sec:energy_functional}

In this section we study the eigenvalue problem
associated with the identity in Eq.~\eqref{eq:H_laplacian_shifted},
first from the combinatorial side. We show that \mct on the HG,
relaxed from the sign variables to the unit sphere, becomes the
eigenvalue problem for the Laplacian, so that computing the GS energy
is solving a spectrally relaxed \mct instance. This also shows the
origin of the edgewise antiferromagnetic condition. We then look at
the same problem from the side of the Hamiltonian and its GS, and
identify the optimal sign structure of the variational wavefunction as the
manifestation of that condition. Finally, at fixed amplitudes, we study the complexity of phase reconstruction.

\subsection{Spectral Relaxation of HG \mct}\label{sec:spectral_relaxation}

Equation~\eqref{eq:H_laplacian_shifted} $H_\ell=E_F-L_\ell$ says that the GS of $H_\ell$ is the  eigenvector of $L_\ell$ with the largest eigenvalue,
\[
E_{\ell, 0} = E_F - \lambda_\mathrm{max} (L_\ell).
\]
Since $L_\ell=D_\ell-W_\ell$, with each diagonal entry of $D_\ell$ equal to the sum of the weights in the corresponding row of $W_\ell$, the expectation value of $L_\ell$ for a real vector $|x\rangle$ with components
$x_\sigma=\langle\sigma|x\rangle$ is a sum over HG edges. Expanding $|x\rangle$ in the orthonormal eigenbasis of the real symmetric matrix $L_\ell$ bounds it from above:
\be\label{variational principle laplacian}
\frac{\braket{x | L_\ell | x}}{\braket{x|x}} = \sum_{\{\sigma, \tau\} \in \mc E_\ell} (W_\ell)_{\sigma \tau} \frac{(x_\sigma - x_\tau)^2}{\braket{x | x}}\leq \lambda_\mathrm{max}(L_\ell).
\ee
Here, we used the fact that each row of the Laplacian $L_\ell$ sums to zero. This inequality is the variational principle expressed in terms of the graph Laplacian, and equality holds only when $\ket{x}$ is the largest eigenvector of the Laplacian.  Additionally, this inequality tells us that the GS is the real vector whose components vary as strongly as possible across HFs, in the weighted quadratic sense of Eq.~\eqref{variational principle laplacian} at fixed norm. We immediately see that to maximize $\braket{x | L_\ell |x}$, we should prioritize opposite signs for $x_\sigma$ and $x_\tau$, since this increases $(x_\sigma - x_\tau)^2$. In other words, we must try to satisfy the condition,
\be\label{pec 0}
\mathrm{sign}(x_\sigma) \  \mathrm{sign}(x_\tau) = -1,
\ee
for as many edges in HG as possible. Setting this condition for all edges is possible if and only if $\Gamma_\ell$ is bipartite. In the presence of frustration or odd cycles, we cannot satisfy this condition for all edges, and in that case we have to prioritize some edges over others depending on the values of the weights and values of $|x_\sigma|, \, | x_\tau|$. This is the origin of \prp for GS. The frustration in the sign structure of the GS is directly induced by the odd cycles in HG. Theorem~\ref{thm:bipartite_inheritance} states that odd cycles in HG directly follow from odd cycles in the physical graph. Thus, the physical graph topology dictates the frustration in the GS sign structure, and choosing the optimal sign structure becomes a combinatorial optimization problem over HG. In fact, this relates to \mct on HG, as we show in the current section and the next.

In a graph, an edge is said to be cut when the two vertices end up on different sides of the cut. We use $\mathrm{sign}(x_\sigma)$ to show which side of the cut the vertex $\sigma$ is on. Hence, the edge $\{\sigma, \tau\}$ is a cut edge if $x_\sigma$ and $x_\tau$ have opposite signs; otherwise, the edge is not cut. A weight $(W_\ell)_{\sigma \tau}$ is associated with each edge. The weighted cut value, determined by the sign pattern of the vector $\ket{x}$, is the sum of the weights of all cut edges.
\be\label{def: cut coupling weighted}
\mathrm{Cut}_{W_\ell}(x) = \sum_{\{\sigma, \tau\} \in \mc E_\ell} (W_\ell)_{\sigma \tau} \,
\frac{1 - \mathrm{sign}(x_\sigma) \, \mathrm{sign}(x_\tau)}{2}.
\ee
Restricting to binary values, $x_\sigma = \pm 1$, then $(x_\sigma - x_\tau)^2 = 2 \, (1 - x_\sigma x_\tau)$ and the cut value can be written in terms of the expectation of the Laplacian
\[
\mathrm{Cut}_{W_\ell}(x) = \frac{1}{4} \, \braket{x | L_\ell | x} ; \quad \ket{x} \in \{\pm 1\}^{|\mc V_\ell|}.
\]
Geometrically, this restriction to binary values means we choose points from the vertices of a Boolean hypercube, as shown in Fig.~\ref{fig:spectral_relaxation}. Now, let $\mc C_\ell$ denote the vertex set of the Boolean hypercube, and let $\mc{S}_\ell$ denote the relaxation of the cube to the hypersphere of radius $\sqrt{|\mc V_\ell|}$
\begin{align*}
\mc S_\ell &= \{\ket x \in \mathbb R^{|\mc V_\ell|} : \braket{x | x} = | \mc V_\ell |\} \\
\mc C_\ell &= \{\ket x \in \mc S_\ell : x_\sigma = \pm 1 \}.
\end{align*}
Then, determining the \mct value and the maximum eigenvalue of the Laplacian becomes the following objectives over these sets
\begin{align*}
\mct(W_\ell) &=
\max_{\ket{x} \in \mc C_\ell} \frac{1}{4} \, \braket{x | L_\ell | x} \\
\lambda_\mathrm{max}(L_\ell) &=
\max_{\ket{x} \in \mc S_\ell} \frac{1}{|\mc V_\ell|} \, \braket{x | L_\ell | x}.
\end{align*}
Moreover, we can extend this result to the complex hypersphere because $L_\ell$ is Hermitian.
\begin{align*}
\mc S^\star_\ell &= \{\ket x \in \mathbb C^{|\mc V_\ell|} : \braket{x | x} = | \mc V_\ell |\}  \\
\lambda_\mathrm{max}(L_\ell) &=
\max_{\ket{x} \in \mc S^\star_\ell} \frac{1}{|\mc V_\ell|} \, \braket{x | L_\ell | x}.
\end{align*}
These equations connect \mct to the largest eigenvalue of the graph Laplacian~\cite{Mohar1990}. Relaxing the search space of \mct from the vertices of the Boolean hypercube to the corresponding real or complex hypersphere yields the graph Laplacian's maximum eigenvalue. In other words, searching for the GS of $H_\ell$ can be interpreted as the spectral relaxation of \mct on HG, as described in Fig.~\ref{fig:spectral_relaxation}. Using~\eqref{variational principle laplacian} we can get an inequality relation between the discrete optimization (\mct) and its relaxed continuous counterpart ($\lambda_\mathrm{max}$).
\[
\mct(W_\ell) \leq \frac{|\mc V_\ell|}{4} \, \lambda_\mathrm{max} (L_\ell)
\]
In terms of GS energy,
\[
E_{\ell, 0} \leq E_F - \frac{4}{|\mc V_\ell|} \mct(W_\ell).
\]
The spectral relaxation above therefore acts on the bare coupling-weighted HG
and allows the vector $\ket{x}$ to vary over the full sphere
$\mc{S}_\ell$, thereby optimizing its component magnitudes and
signs simultaneously (Fig.~\ref{fig:HG}). To isolate phase reconstruction, we now resolve each variational coefficient into a positive amplitude and a phase. Once the amplitude profile is fixed, only the phase degrees of freedom, or equivalently the sign degrees of freedom in the real sector, remain, and the bare HG edge weights are dressed by amplitude products. The resulting discrete sign optimization problem is a weighted \mct problem on the same HG, but it is generally not the bare \mct instance considered above.

\subsection{The Phase Reconstruction Problem}\label{sec:energy_prp}

We now study how the wavefunction's sign structure enters the energy minimization. Eq.~\eqref{pec 0} already showed that the Laplacian favors opposite signs on neighboring HG vertices for its largest eigenvalue. To make this dependence explicit, we separate the wavefunction into its amplitude and phase. We write a variational state in the sector $\ell$ as
\be\label{eq:variational_state}
\ket{\Psi_\ell} = \sum_{\sigma \in \mc V_\ell} \psi_\sigma e^{i \phi_\sigma} \ket{\sigma}; \quad
\psi_\sigma, \phi_\sigma \in \mathbb R, \, \psi_\sigma \geq 0
\ee
and the energy associated with this state is
\[
E_\ell = \frac{\braket{\Psi_\ell | H_\ell | \Psi_\ell}}{\braket{\Psi_\ell | \Psi_\ell}},
\]
when the variational state is non-zero. Since no two sectors share any basis states, we can write any variational state in the full computational basis as,
\[
\ket{\Psi} = \sum_{\ell = 0}^{N} \ket{\Psi_\ell}
\]
Our focus in this section is the phase reconstruction problem or \prp. We define \prp as solving for the phases $\phi_\sigma$ such that the energy expectation of the variational state is minimized for given amplitudes $\psi_\sigma$.
\begin{definition}[Phase Reconstruction Problem (\prp)]
Let $\{\ket{\sigma}\}$ be an orthonormal basis of the Hilbert space, let $\hat H$ be the Hamiltonian, and let $\{\psi_\sigma\}$ be a fixed set of non-negative amplitudes. The phase reconstruction problem consists of finding the phases $\phi_\sigma \in \mathbb{R}$ that minimize the variational energy,
$\braket{\Psi|\hat H|\Psi}$ for $\ket\Psi = \sum_\sigma \psi_\sigma e^{i\phi_\sigma}\ket\sigma$. Since wavefunctions have $U(1)$ symmetry, the solution is defined up to a global phase.
\end{definition}

For a non-zero state, the energy expectation in the full computational basis reduces to:
\[
E = \frac{\braket{\Psi | \hat H | \Psi}}{\braket{\Psi | \Psi}}
= \sum_{\braket{\Psi_\ell | \Psi_\ell} \neq 0} \frac{\braket{\Psi_\ell | \Psi_\ell}}{\braket{\Psi | \Psi}} \, E_\ell
\]
Since $\braket{\Psi_\ell | \Psi_\ell}$ are independent of phase, separately minimizing each $E_\ell$ will also minimize $E$. In other words, solving \prp in each sector separately solves \prp for the entire Hilbert space. Accordingly, our objective now shifts to energy minimization in each sector separately.

The Hamiltonian in each sector separates into classical and quantum parts, $H_\ell = H_{\ell,c} + H_{\ell,q}$, and so does the energy, $E_\ell = E_{\ell,c} + E_{\ell,q}$.
\begin{align*}
E_{\ell,c} = \frac{\braket{\Psi_\ell | H_{\ell, c} | \Psi_\ell}}{\braket{\Psi_\ell | \Psi_\ell}},
&&E_{\ell,q} = \frac{\braket{\Psi_\ell | H_{\ell,q} | \Psi_\ell}}{\braket{\Psi_\ell | \Psi_\ell}}
\end{align*}
The classical part $E_{\ell,c}$ is diagonal and thus independent of phase. For phase optimization, we therefore only need to pay attention to the quantum part $E_{\ell,q}$, which contains interference terms when written as a sum over HG edges.
\be\label{eq:5}
E_{\ell, q} = \sum_{\{\sigma, \tau\} \in \mc E_\ell} (w_\ell)_{\sigma \tau} \, \cos(\phi_\sigma - \phi_\tau),
\quad (w_\ell)_{\sigma \tau} \geq 0.
\ee
Here $w_\ell$ is the sector $\ell$ amplitude-dressed weight matrix.
\be\label{def: amplitude weighted adjacency}
(w_\ell)_{\sigma \tau} = \frac{2 \, \psi_\sigma \psi_\tau}{\braket{\Psi_\ell | \Psi_\ell}}(H_{\ell, q})_{\sigma \tau}
\ee
Eq.~\eqref{eq:5} is an amplitude-weighted antiferromagnetic XY model on the HG, where each edge in HG has an energy $(w_\ell)_{\sigma\tau} \cos(\phi_\sigma - \phi_\tau)$ associated with it. For a physical lattice $G$ and coupling ratio $J_2/J_1$, the coupling-dressed weights are fixed. However, the amplitude-dressed weights are \textit{emergent}: the Hamiltonian provides the bare coupling $J_r$, while the variational state dresses each edge by $\psi_\sigma\psi_\tau/\braket{\Psi_\ell|\Psi_\ell}$. As the amplitudes vary, the same HG therefore supports a family of amplitude-weighted XY models in each sector.

Since the active edge weights in Eq.~\eqref{eq:5} are nonnegative, the energy minimization is governed by the interference term $\cos(\phi_\sigma-\phi_\tau)$.
Every edge weight $(w_\ell)_{\sigma\tau}$ is non-negative for antiferromagnetic couplings, i.e. $J_r\geq0$, and the edge energy is minimized by $\cos(\phi_\sigma-\phi_\tau)=-1$, which we call the $\pi$-edge condition (PEC). The opposite case, $\cos(\phi_\sigma-\phi_\tau)=+1$, is the zero-edge condition
(ZEC).
\begin{align}
\text{(PEC)}\quad \phi_\sigma-\phi_\tau &\equiv \pi \pmod{2\pi}, \label{eq:7}\\
\text{(ZEC)}\quad \phi_\sigma-\phi_\tau &\equiv 0 \pmod{2\pi}. \label{eq:zec}
\end{align}
Every edge favors PEC as it locally minimizes the energy, and this preference is independent of the amplitudes: the weights set the magnitude of each edge contribution but never its sign. Whether all edges can satisfy PEC simultaneously therefore depends on the HG's topology.

Now, consider the assignments of phase factor $g_{\sigma\tau}\in U(1)$ to every edge of $\Gamma_{\ell}(G)$ with $g_{\sigma\tau}^{-1}=g_{\tau\sigma}$. A phase field $\{\phi_\sigma\}$ induces such an assignment through
\be\label{eq:phase_factor}
g_{\sigma\tau} := e^{i(\phi_\sigma-\phi_\tau)},
\ee
but not every assignment is of this form. For any closed loop $\gamma = (\sigma_0,\sigma_1,\dots,\sigma_{n-1},\sigma_n=\sigma_0)$ in $\Gamma_{\ell}(G)$, the \textit{holonomy} of the phase field around $\gamma$ is
\be\label{eq:holonomy_first}
\texttt{Hol}_\gamma := \prod_{k=0}^{n-1} g_{\sigma_k\sigma_{k+1}}.
\ee
It is invariant under $g_{\sigma\tau} \rightarrow h_{\sigma}g_{\sigma\tau}h_{\tau}^{-1}$ for arbitrary $h_{\sigma}\in U(1)$, which is the gauge transformation $\phi_{\sigma} \mapsto \phi_{\sigma} + \alpha_{\sigma}$ with an arbitrary angle $\alpha_{\sigma}$ at each vertex.
Since each $g_{\sigma\tau}$ is a difference of vertex phases, the product telescopes and $\texttt{Hol}_\gamma=1$ for every closed loop, whatever the phase field. A global PEC, on the other hand, prescribes $g_{\sigma\tau}=-1$ on every edge, so $ \texttt{Hol}_\gamma = (-1)^{|\gamma|} $.
The two requirements are compatible only if every closed walk in $\Gamma_{\ell}(G)$ has even length. This leads us to re-express the two-coloring theorem in the context of PEC on HG.
\begin{corollary}[PEC--bipartiteness]\label{cor:PEC_bipartite}
A global $\{0,\pi\}$ phase assignment obeying PEC on every edge exists if and only if $\Gamma_\ell(G)$ is bipartite.
\end{corollary}
The forward proof follows directly from the holonomy argument above. Conversely, if $\Gamma_{\ell}(G)$ is bipartite, assigning phases $0$ and $\pi$ to parts yields PEC on every edge. Since each edge contribution to $E_{\ell,q}$ is bounded below by $-(w_\ell)_{\sigma\tau}$, this assignment saturates the lower bound on every edge simultaneously and therefore gives
\[
E_{\ell, q}=-\sum_{\{\sigma,\tau\}\in\mathcal{E}_\ell} (w_\ell)_{\sigma\tau}.
\]
Corollary~\ref{cor:PEC_bipartite} establishes the Marshall phase pattern as an exact PEC assignment for the bipartite HAF; for a given sublattice $V_A$, an example of this assignment is given by $\phi_\sigma = \pi N_{A}^{\uparrow}(\sigma)$~\eqref{eq:N_A}. The same assignment also minimizes the phase-dependent energy when the NN couplings are antiferromagnetic, and the NNN couplings are ferromagnetic, provided the NNN bonds connect sites within the same physical sublattice. An NN exchange changes the number of up spins on sublattice A by one, giving a phase difference of $\pi$ and satisfying PEC. An NNN exchange leaves this number unchanged, giving zero phase difference and satisfying ZEC. Both contributions therefore attain their lower bounds simultaneously for fixed amplitudes. Thus, the mixed AFM–FM model is sign-unfrustrated, even when its HG is non-bipartite.  We emphasize that Marshall's original proof~\cite{marshall1955} proceeds by contradiction and does \textit{not} use the graph-theoretical viewpoint adopted here.

\begin{figure}[t]
\centering
\includegraphics[width=\columnwidth]{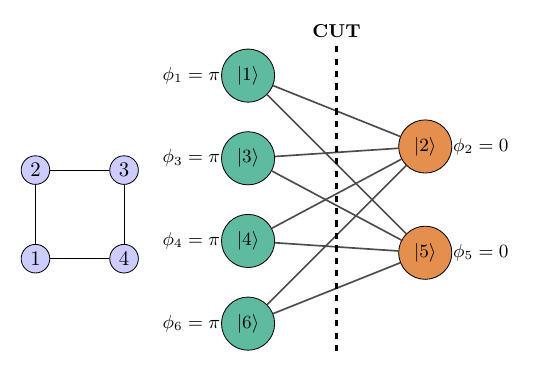}  
\caption{\textbf{Left:} A $2\times2$ open boundary square lattice NN
HAF. \textbf{Right:} The corresponding bipartite HG, $K_{2,4}$, in
the zero magnetization sector; spin configurations as in
Fig.~\ref{fig:HG}. The dashed line is the cut separating the two
N\'eel configurations (orange) from the rest (green). Because the
HG is bipartite, this cut contains every edge: the phase assignment
$\phi_\sigma=\pi$ on one side and $\phi_\sigma=0$ on the other
satisfies PEC on all edges simultaneously, saturating the edgewise
bound and attaining $E_{\ell, q}=-\sum_{\{\sigma,\tau\}\in\mc E_\ell} (w_\ell)_{\sigma\tau}$. The \mct is independent of the edge weights for \textit{any} bipartite HG.}
\label{fig:lattice-and-cg}
\end{figure}

For the bipartite HG, PEC holds globally; hence all edges are cut, as shown in Fig.~\ref{fig:lattice-and-cg}.
An odd cycle cannot fully satisfy the PEC and therefore leads to at least one ZEC edge; the geometric frustration in the HG is thus this topological obstruction to achieving global PEC. Moreover, under frustration, the amplitudes assume a role they did not have before. The set of violated edges is no longer empty, and the best choice of which edges to leave violated depends on the weights $w_{\sigma\tau}$; different amplitude profiles result in different optimal sign structures for the same HG. Consequently, when frustration is present, the solution to \prp depends on the amplitudes. For a given choice of amplitudes, minimizing $E_q$ with respect to assignments in $\{0,\pi\}$ becomes a global optimization problem, equivalent to an \mct problem on the weighted HG, as we shall show below.

\begin{figure*}[htbp]
    \centering
    \includegraphics[width=1.0\textwidth]{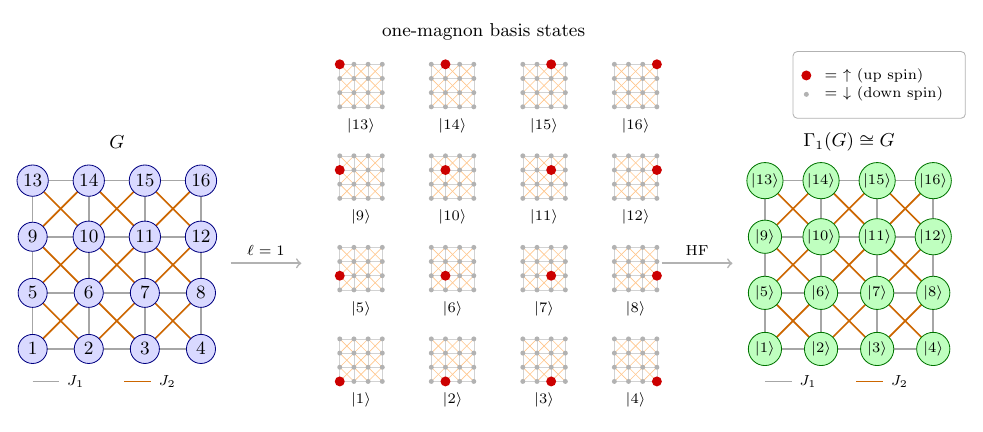}
\caption{One-magnon HG of the $4\times4$ square-lattice $J_1$--$J_2$ HAF with open boundaries. \textbf{(Left)} The physical lattice, with NN bonds $J_1$ shown in gray and NNN bonds $J_2$ in orange. \textbf{(Middle)} The one-magnon basis states. Each state $\ket{i}$ contains one up spin at site $i$, shown in red, with all remaining spins down. \textbf{(Right)} HFs move the up spin along the $J_1$ and $J_2$ bonds, reproducing the physical connectivity giving $\Gamma_1(G)\cong G$.
}
\label{fig:one magnon HG}
\end{figure*}

\subsection{Binary \prp and \mct}\label{sec:binary_prp}
We will now restrict the optimization to real wavefunctions. This is equivalent to optimizing the phases over $\{0, \pi\}$. This restriction is particularly useful for understanding the GS phase structure. The HAF Hamiltonian has real matrix elements; hence, we can choose real eigenvectors without loss of generality. Then, for GS amplitudes, \prp over binary values and \prp over continuous values yield the same answer, since the global minima of $E_q$ lie at a discrete phase assignment for real GS, ignoring the global phase of the wavefunction. We define the signs,
\[
s_\sigma = e^{i \phi_\sigma} \in \{+1, -1\}, \qquad \phi_\sigma \in \{0, \pi\}.
\]
Then $\cos(\phi_\sigma-\phi_\tau)$ can be replaced with $s_\sigma s_\tau$ and
\[
E_{\ell, q} = \sum_{\{\sigma, \tau\} \in \mc E_\ell} (w_\ell)_{\sigma\tau} \, s_\sigma s_\tau,
\qquad (w_\ell)_{\sigma\tau} \geq 0.
\]
The quantum part of the energy functional therefore defines a classical antiferromagnetic Ising model on the HG~\cite{Westerhout:2022lil}. For a fixed amplitude profile, finding the signs that minimize the energy is equivalent to a QUBO instance~\cite{Lucas2014}. When weights are generated from GS amplitudes, the relevant sector is $\ell = N/2$, or the zero-magnetization sector, since this sector is guaranteed to contain a GS for an even number of spins, as shown in Sec.~\ref{subsec:laplacian_spectrum}. 

Since the amplitudes are held fixed for \prp, the edge weights are also fixed. Minimizing the Ising energy is then equivalent to maximizing the corresponding weighted cut. Similar to Sec.~\ref{sec:spectral_relaxation}, any assignment $\{s_\sigma\}$ divides the vertices of $\Gamma_{\ell}(G)$ into
two sets and thus defines a cut. An edge $\{\sigma,\tau\}$ contributes
$-(w_\ell)_{\sigma\tau}$ to $E_{\ell, q}$, if $\sigma$ and $\tau$ lie on opposite sides of the cut, i.e. $s_\sigma s_\tau=-1$, and $+(w_\ell)_{\sigma\tau}$ if they lie on the same side, i.e. $s_\sigma s_\tau=+1$. Similar to Eq.~\eqref{def: cut coupling weighted} we define the cut value in this case as,
\be\label{def: cut amplitude weighted}
\mathrm{Cut}_{w_\ell}(s) = \sum_{\{\sigma, \tau\} \in \mc E_\ell} (w_\ell)_{\sigma \tau} \,
\frac{1 - s_\sigma \, s_\tau}{2},
\ee
and the energy can then be written as,
\be\label{eq:fixed_amplitude_prp3}
E_{\ell, q} (s) = \sum_{\{\sigma,\tau\} \in \mc E_\ell} (w_\ell)_{\sigma\tau} - 2 \, \mathrm{Cut}_{w_\ell} (s).
\ee
Since the first term in Eq.~\eqref{eq:fixed_amplitude_prp3} is constant, minimizing $E_{\ell, q}$ is equivalent to maximizing the cut value $\mathrm{Cut}_{w_\ell}(s)$. Binary \prp is therefore an instance of weighted \mct on the HG, leading to the following theorem.
\begin{theorem}[Heisenberg-Ising Correspondence]
Binary or $\{0, \pi\}$ \prp of HAF defined on a physical graph G in sector $\ell$, is equivalent to determining the GS of a classical antiferromagnetic Ising model on $\Gamma_\ell(G)$, which in turn is equivalent to a weighted \mct problem on $\Gamma_\ell(G)$.
\end{theorem}

In this correspondence, the Ising interaction on the HG arises from the interference between two basis configurations connected by the off-diagonal Heisenberg exchange: in the binary sector, the relative phase of this interfering pair is encoded by the product of two sign variables attached to the corresponding HG vertices. Consequently, finding the optimal binary sign structure is the GS problem of this weighted antiferromagnetic Ising model, or equivalently, the weighted \mct problem on the amplitude-weighted HG, illustrated in Fig.~\ref{fig:HG}.

Each binary \prp of HAF maps, in polynomial time, to a \mct instance on $\Gamma_\ell(G)$ carrying the weights induced by the amplitudes through Eq.~\eqref{def: amplitude weighted adjacency}. This polynomial-time reduction can be written as 
\be\label{eq:complexity_statement}
\textsf{Binary~}\textsf{PRP}\le_{\mathrm{p}}\textsf{Max-Cut},
\ee
meaning that any polynomial-time algorithm for weighted \mct would immediately yield one for binary \prp. The relation~\eqref{eq:complexity_statement} bounds the difficulty of \prp only from above: it shows that \prp is at most as hard as weighted \mct, but does not, by itself, establish \np-hardness.

A true hardness statement would require the reverse reduction,
$\textsf{Max-Cut}\le_{\mathrm p}\textsf{Binary }\prp$, namely a polynomial-time encoding of an arbitrary \mct instance into a \prp instance. The HG of a HAF is a token graph; hence, it has a very specific structure. So it is not obvious that for \mct defined any graph we can find an isomorphic HG. Moreover, the weights come from the amplitudes and thus have a specific structure.

Nevertheless, we can map \mct on arbitrary graph to binary \prp on an HG by using the one-magnon sector, or $\ell=1$. This happens because in the one-magnon sector, the HG is isomorphic to its parent graph (Fig.~\ref{fig:one magnon HG}). If we use the Heisenberg model of Eq.~\ref{def: general Heisenberg Model} then for an $N$-spin system, fixing $\psi_\sigma = 1 \, \forall \sigma \in \mc V_1$, it follows directly from Eq.~\eqref{def: amplitude weighted adjacency} that
\be\label{eq: weights in one magnon}
(w_1)_{\sigma\tau} = J_{ij} / N \quad \forall \sigma \overset{ij}{\rightleftharpoons} \tau.
\ee
Thus, the parent graph weights get inherited by its HG. This implies that, by setting the right coupling values, we can map any \mct in $G$ to an equal-amplitude binary \prp in $\Gamma_1(G)$. By the duality~\eqref{eq:duality} the same argument also covers $\ell=N-1$.  Ref.~\cite {ruy_inprep} we classifies the complexity of unweighted \mct on HG as a function of $\ell$. Notable results are: for a fixed constant value of $\ell$, unweighted \mct on HG is \np-hard in the size of the physical graph, and for $\ell \ge N^\delta, \delta > 0$, unweighted \mct on HG is not \np-hard in the size of the HG if a quasipolynomial counterpart of the hypothesis $\np \subseteq \mathrm{P}/\mathrm{poly}$ holds true.

\section{Optimization Methods}\label{sec:relaxation}
Many algorithms have been developed for solving \mct. Here we focus on two that are intricately tied to studying \mct on HG: the \rt relaxation and the \gw algorithm. The \rt relaxation is already present in the quantum mechanical expectation value of the energy and the \gw algorithm is capable of producing exact GS sign structure in time polynomial in the size of the HG, as we will discuss in the following sections.

\subsection{\rt Relaxation}\label{sec: rank-2 relaxation}
\begin{figure*}[htbp]
\centering
\includegraphics[width=1.00\textwidth]{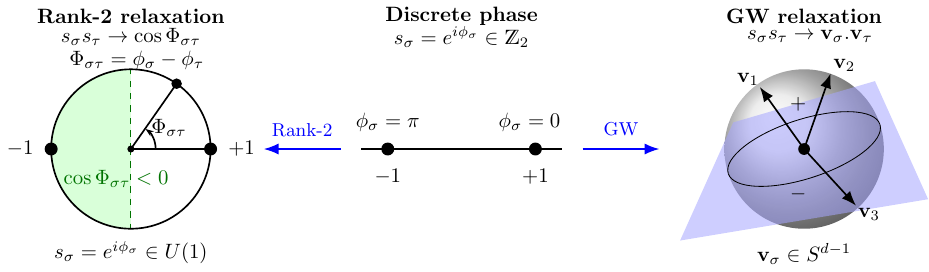}
\caption{
Comparison of discrete phases, \rt relaxation, and the Goemans-Williamson relaxation. \textbf{Center:} the original discrete phase variable is restricted to the Ising values $s_\sigma=e^{i\phi_\sigma}\in\mathbb{Z}_2$, with $\phi_\sigma=0$ or $\pi$. \textbf{Left:} the \rt relaxation promotes the sign variable to a continuous phase $s_\sigma=e^{i\phi_\sigma}\in U(1)$, so that the discrete product $s_\sigma s_\tau$ is relaxed to the continuous edge correlation $\cos\Phi_{\sigma\tau}$, where $\Phi_{\sigma\tau}=\phi_\sigma-\phi_\tau$. The shaded half disk marks phase differences $\Phi_{\sigma\tau}$ (not absolute phases) satisfying $\cos\Phi_{\sigma\tau}<0$, which lower the off-diagonal energy contribution $E_q$ defined in Eq.~\eqref{eq:5}. \textbf{Right:} the Goemans-Williamson relaxation lifts the Ising variable to a unit vector
$\mathbf v_\sigma\in S^{d-1}$ and replaces $s_\sigma s_\tau$ by $\mathbf v_\sigma\cdot\mathbf v_\tau$. The hyperplane separates the sphere into two hemispheres, corresponding to the two signs after rounding.
}
\label{fig:rank_two_phase_relaxation}
\end{figure*}

In this section we study \prp without restricting the phases. Here, we minimize the weighted XY energy $E_q(\phi)$ of Eq.~\eqref{eq:5} on the HG with fixed weights by continuously optimizing the phases. The phases form points on a circle, as shown in Fig.~\ref{fig:rank_two_phase_relaxation}. This optimization is similar to the \rt relaxation method of \textcite{burer2002rank} for solving \mct. \rt relaxation was introduced as a cheap method for approximating \mct of weighted graphs. We, however, arrived at this method from a purely quantum-mechanical perspective; the complex wavefunction and the Hamiltonian's Hermiticity force the \rt form of Eq.~\eqref{eq:5}. But \rt relaxation is defined for any weighted graph, whereas \prp is defined on weighted HG. In the one-magnon sector, however, \prp and \rt relaxation are equivalent. This is because $\Gamma_1(G) \cong G$ and Eq.~\eqref{eq: weights in one magnon} implies that for equal amplitude \prp in one magnon sector one can always choose couplings of the Hamiltonian~\eqref{def: general Heisenberg Model} such that arbitrary weighted \mct on $G$ can also be realized on $\Gamma_1(G)$.

The phase-dependent part of any variational ansatz in sector $\ell$ with continuously parameterized phases is $E_{\ell,q}$. However, not all variational ansätze allow independent changes to amplitudes (hence weights) and phases. Only mod-phase networks allow that~\cite{szabo}. Then, for these networks, using the \rt relaxation formulation, we can extract information about the energy landscape in the phase sector, treating phase and amplitude as independent parameters. We discuss this in detail later; for now, we introduce basic properties of \rt relaxation. We ask what the landscape of $E_q(\phi)$ looks like, namely its stationary points, their character, and how the discrete $\{0, \pi\}$ assignments sit inside the continuum. The objective of \rt relaxation is minimization of
\be\label{eq: rank-2 objective}
\min_{\phi_\sigma \in \mathbb R} E_{\ell,q}(\phi) = \min_{\phi_\sigma \in \mathbb R} \sum_{\{\sigma, \tau\} \in \mc E_\ell} (w_\ell)_{\sigma \tau} \cos(\phi_\sigma - \phi_\tau),
\ee
where the weights are fixed. The problem has a $U(1)$ symmetry; shifting all $\phi_\sigma$ by a global phase does not change the energy. \rt relaxation first finds minima of $E_{\ell,q}$ using gradient descent methods, then applies a rounding method to convert phases to binary values, $\pm 1$. The minimization part is \prp and the rounding part is binary \prp. The details of the algorithm are given in Algorithm~\ref{alg: rank-2}. The gradient of $E_{\ell, q}$ with respect to the phase is given by,
\be\label{eq: dE_q/dphi}
\frac{\partial E_{\ell,q}}{\partial \phi_\sigma} = 
- \sum_{\tau\in\mathcal N(\sigma)} (w_\ell)_{\sigma\tau} \, \sin(\phi_\sigma-\phi_\tau),
\ee
Minimizing $E_{\ell, q}$ with respect to the phases $\{\phi_{\sigma}\}$ yields the Karush--Kuhn--Tucker (KKT)~\cite{karush1939nonlinear,kuhn1951nonlinear} stationary conditions
\be\label{eq:phase_balance}
\sum_{\tau\in\mathcal N(\sigma)}
(w_\ell)_{\sigma\tau} \,
\sin(\phi_\sigma-\phi_\tau)=0,
\qquad \forall\,\sigma \in \mc V_\ell.
\ee
The solutions of these equations correspond to zero gradients of $E_{\ell,q}$ with respect to $\phi_\sigma$. The discrete phase assignment $\phi_\sigma\in\{0,\pi\}$ (modulo $2\pi$) is an obvious solution of the stationarity equations. For $n$ states, there are $2^n$ such discrete phase assignments. Whenever a variational ansatz can realize such an assignment, these solutions also correspond to zero gradients of $E_{\ell,q}$ with respect to the ansatz's variational parameters, provided the amplitudes are held fixed, by the chain rule. Moreover, if $\{\phi_\sigma\}$ is a solution, then $\{\phi_\sigma+\delta\}$ is also a solution for any constant $\delta$. As a result, these $\{0,\pi\}$ assignments organize into $2^{n-1}$ distinct one-parameter families (``lines'') in the phase space (modulo $2\pi$) that correspond to stationary points of $E_{\ell, q}$. In general, there can be stationary points that do not correspond to a $\{0, \pi\}$ assignment.

The local character of the stationary points is determined by the Hessian
\[
\frac{\partial^2 E_{\ell,q}}{\partial \phi_\sigma \partial \phi_\tau}
=
\begin{cases}
-\sum_{\mu\in\mathcal N(\sigma)}
(w_\ell)_{\sigma\mu}\cos(\phi_\sigma-\phi_\mu),
& \sigma=\tau,\\
\displaystyle (w_\ell)_{\sigma\tau}\cos(\phi_\sigma-\phi_\tau),
& \{\sigma,\tau\}\in\mathcal E_\ell,\\
0, & \text{otherwise}.
\end{cases}
\]
If $(w_\ell)_{\sigma\tau}\cos(\phi_\sigma-\phi_\tau)\ge 0$ on every edge, then the Hessian is negative semidefinite due to diagonal dominance and the Gershgorin disk theorem; accordingly, if
$(w_\ell)_{\sigma\tau}\cos(\phi_\sigma-\phi_\tau)\le 0$ on every edge, then it is positive semidefinite. In general, however, these effective couplings have mixed signs, so the Hessian need not be semidefinite and is generically indefinite. In fact, if a $\{0, \pi\}$ assignment does not correspond to a global minimum or maximum, then it is a saddle point (see Theorem 3.4 of Ref.~\cite{burer2002rank}).

\begin{algorithm}[h]
\caption{\rt \mct}
\label{alg: rank-2}
\begin{algorithmic}[1]
\Require weight matrix $w_{ij}$ and indices $i, j$; number of vertices $n$; steps $T$; Adam~\cite{Adam} optimizer.
\State Normalize the weights, $w_{ij} \gets w_{ij} / \mathrm{max}(|w_{ij}|)$.
\State Draw angles $\theta_i \in \mathbb{R}$ uniformly from $[0, \pi)$, so that $\mathbf v_i = (\cos\theta_i, \sin\theta_i)$.
\For{$t=1,\dots,T$}
  \State Calculate gradient, $G_i \gets \sum_{j} (w_{ij} + w_{ji}) \, \sin(\theta_j - \theta_i)$.
  \State Update angles using optimizer, $\theta_i \gets \theta_i + \mathrm{Adam}(G_i)$.
\EndFor
\State Map the angles to the circle, $\theta_i \gets \theta_i \bmod 2\pi$.
\State Sort $\{\beta_i = \theta_i \bmod \pi\}$ in ascending order as $\beta_1 \le \dots \le \beta_n$, and set $\beta_{n+1} = \pi$.
\State Construct the candidate cut angles, $\alpha_m \gets (\beta_m + \beta_{m+1})/2$ for $m = 1,\dots,n$.
\State Construct the sign matrix, $S_{mi} = +1$ if $\theta_i \le \alpha_m$ or $\theta_i > \alpha_m + \pi$, and $S_{mi} = -1$ otherwise.
\State Choose $m = m^\star$ such that $\sum_{ij} w_{ij} \, S_{mi} S_{mj}$ is minimum.
\State Return the sign structure, $\{s_i\}$ for $s_i = S_{m^\star i}$.
\end{algorithmic}
\end{algorithm}

The \rt objective is non-convex, and spurious minima can trap the optimization at higher energies. \textcite{burer2002rank} mitigate this by random restarts where after reaching a local minimum the phases are moved to the lowest energy $\{0, \pi\}$ assignment in its neighborhood, which is a saddle point, and the minimization is restarted from there in hope of reaching a lower minimum. But even if the global minimum is reached this way, it may not correspond to a cut, i.e., $\{0,\pi\}$ phase values. However, since the eigenvectors of HAF can be chosen real, when the weights are generated from GS amplitudes, the global minima are situated at a $\{0, \pi\}$ assignment (modulo $2\pi$). Therefore, the accuracy of GS \prp using \rt relaxation is only limited by the quality of the gradient descent. Table~\ref{tab:gw-vs-rank2} shows the accuracy of \rt relaxation for GS weights. Here, the Adam~\cite{Adam} optimizer was used for gradient descent without random restarts, and the results in Table~\ref{tab:gw-vs-rank2} imply that the optimizer was able to reach very close to the global minima.

The \rt relaxation maps discrete binary-valued signs to points on a unit circle. One can also map the signs to a higher-dimensional sphere, which leads to the next section.

\subsection{Goemans--Williamson Relaxation}\label{sec:gw}

The Goemans--Williamson (\gw) algorithm~\cite{goemans1995improved} replaces the Ising spins $s_{\sigma} \in \{\pm 1 \}$ of Eq.~\eqref{def: cut amplitude weighted} with unit vectors $\mathbf{v}_{\sigma} \in \mathbb{R}^{|\mathcal V_\ell|}$. Using $\mathbf{v}_{\sigma} \cdot \mathbf{v}_{\tau}$ as a continuous proxy for $s_{\sigma}s_{\tau}$, it relaxes \mct to
\be\label{eq:gw2}
\max_{\| \mathbf v_\sigma \| = 1} 
\sum_{\{\sigma,\tau\}\in \mc{E}_\ell} (w_\ell)_{\sigma\tau} \left( \frac{1 - \mathbf{v}_{\sigma} \cdot \mathbf{v}_{\tau}}{2} \right)
\ee
Equivalently, define the Gram matrix $X \in \mathbb{R}^{|\mathcal V_\ell|\times|\mathcal V_\ell|}$ by 
\be\label{eq:gram_matrix}
X_{\sigma\tau} = \mathbf{v}_\sigma \cdot \mathbf{v}_\tau, \qquad \forall \sigma,\tau \in \mc V_\ell
\ee
so that $X = Y^\top Y$, where $Y \in \mathbb{R}^{|\mathcal V_\ell| \times |\mathcal V_\ell|}$ is the matrix whose $\sigma$-th column is $\mathbf{v}_\sigma$. By construction, $X \succeq 0$ and $\mathrm{diag}(X)=I$. The relaxed
\mct then becomes the following SDP.
\be\label{eq:gw3}
\max_{X \succeq 0, \, X_{\sigma\sigma} = 1}
\sum_{\{\sigma,\tau\}\in \mc E_\ell} (w_\ell)_{\sigma\tau} \left( \frac{1 - X_{\sigma\tau}}{2} \right)
\ee
Any sign assignment $\{s_\sigma\}$ can be realized by this relaxation using $\mathbf v_\sigma = s_\sigma \mathbf u$, where $\hat u$ is a fixed unit vector. Thus, the maximum of the relaxed problem is at-least as large as the \mct. This convex program can be solved in polynomial time to obtain an optimal solution $X^\star$, which can be factorized, for example via Cholesky or eigen decomposition, as $X^\star = (Y^\star)^\top Y^\star$, yielding a set of unit vectors $\{\mathbf v^\star_\sigma\}$.

The SDP solution thus gives a set of unit vectors $\{ \mathbf v^\star_{\sigma} \}$  instead of the Ising spins $\pm 1$. \gw converts the vectors into Ising spins using a random hyperplane through the origin: (i) Sample a random direction $\mathbf r$ uniformly from the unit sphere. (ii) Assign a spin by the sign of the projection:
\[
s_{\sigma} = \mathrm{sign}(\mathbf r \cdot \mathbf{v}^\star_{\sigma})
\]
Geometrically, the hyperplane perpendicular to $\mathbf r$ separates the hypersphere into two hemispheres; vertices of an edge landing on opposite hemispheres form a cut. For an edge $\{\sigma, \tau\}$, if the angle between $v_{\sigma}$ and $v_{\tau}$ is $\theta_{\sigma\tau} \in [0,\pi]$, then the probability that the hyperplane separates them is $\mathrm{Pr}[s_{\sigma} \neq s_{\tau}] = \tfrac{\theta_{\sigma\tau}}{\pi}$. Hence, the expected rounded cut value is
\be\label{eq:gw4}
\mathbb{E}[\textsf{Cut}(s)] = \sum_{(\sigma,\tau) \in \mc{E}_\ell} (w_\ell)_{\sigma\tau} \frac{\theta_{\sigma\tau}}{\pi}
\ee
Goemans and Williamson showed that for all $\theta_{\sigma\tau}$,
\be\label{eq:gw5}
\frac{\theta_{\sigma\tau}/\pi}{(1-\cos\theta_{\sigma\tau})/2} \geq \alpha_\gw,
\ee
where $\alpha_\gw \approx 0.87856$. This bound leads to the following relation for non-negative edge weights, i.e., $(w_\ell)_{\sigma\tau} \geq 0$.
\[
\mathbb{E}[\textsf{Cut}(s)] \geq \alpha_\gw
\sum_{\{\sigma,\tau\}\in \mc E_\ell} (w_\ell)_{\sigma\tau} \left( \frac{1 - X^\star_{\sigma\tau}}{2} \right)
\]
Thus, the random-hyperplane rounding produces a cut whose expected value is at least $87.8\%$ of \mct.
The \gw algorithm solves an SDP with a symmetric $|\mc V_\ell|\times|\mathcal V_\ell|$ positive-semidefinite matrix. This means the algorithm deals with $\mc O(|\mathcal V_\ell|^2)$ variational parameters. In contrast, \rt relaxation uses $\mc O(|\mc V_\ell|)$ parameters but has a non-convex optimization landscape. Therefore, we must choose between computational efficiency and theoretical performance guarantees. However, Table~\ref{tab:gw-vs-rank2} shows that for GS \prp, the \rt relaxation can produce high-quality solutions despite being non-convex. Ref.~\cite{burer2002rank} has also found that \rt can produce high-quality approximate solutions while being significantly faster than the native \gw implementation. Moreover, the size of HG grows exponentially when $\ell$ is linear in the number of spins, which means the native \gw implementation becomes memory-constrained for large system sizes much faster than \rt.

For these reasons, we have chosen the Burer-Monteiro (BM) factorization \cite{Burer2003, Burer2005} as a cheap method to implement the \gw algorithm. The BM factorization involves optimization over a $k \times |\mc V_\ell|$ matrix, $Y$, for $X = Y^T Y$. In other words, the BM factorization assumes that it is optimal for the vectors $\mathbf v^\star_\sigma$ to be $k$-dimensional. $X$ is positive semidefinite by construction in this case, and $X_{\sigma\sigma}=1$ is achieved by forcing the vectors to be of unit length. However, when optimizing over $Y$ instead of $X$, we lose convexity, and second-order critical points appear. \textcite{boumal} have shown that as long as $k \geq \lceil \sqrt{2|\mc V_\ell|} \, \rceil$, the second order critical points are also global minima and \textcite{waldspurger} have shown that this bound is tight. Thus we can effectively reduce the number of variables to $\mc O(|\mc V_\ell|^{1.5})$ while retaining the performance guarantee of the \gw algorithm. The implementation details for BM factorization are given in Algorithm~\ref{alg: gw}. The results in Table~\ref{tab:gw-vs-rank2} imply that the optimizer was able to reproduce GS signs faithfully.

\begin{algorithm}[h]
\caption{\gw algorithm via BM factorization}
\label{alg: gw}
\begin{algorithmic}[1]
\Require weight matrix $w_{ij}$ and indices $i, j$; number of vertices $n$, rank $k$; hyperplanes $n_h$; steps $T$; Adam~\cite{Adam} optimizer.
\State Normalize the weights, $w_{ij} \gets w_{ij} / \mathrm{max}(|w_{ij}|)$.
\State Draw $V \in \mathbb{R}^{k \times n}$ from a random distribution.
\State Normalize the column vectors of $V$, $\mathbf v_i \gets \mathbf v_i / |\mathbf v_i| $.
\For{$t=1,\dots,T$}
  \State Calculate gradient, $\mathbf G_i \gets \sum_{j} (w_{ij} + w_{ji}) \, \mathbf v_j$.
  \State Discard the radial part, $\mathbf G_i\gets \mathbf G_i - ( \mathbf G_i \cdot \mathbf v_i ) \, \mathbf v_i$.
  \State Update vectors, $\mathbf v_i \gets \mathbf v_i + \mathrm{Adam}(\mathbf G_i)$.
  \State Normalize the vectors after update, $\mathbf v_i \gets \mathbf v_i / |\mathbf v_i|$.
\EndFor
\State Draw $n_h$, $k$-dimensional random vectors, $\{\mathbf r_m\}$.
\State Construct the sign matrix $S_{mi} = \mathrm{sign}(\mathbf r_m \cdot \mathbf v_i)$.
\State Choose $m = m^\star$ such that $\sum_{ij} w_{ij} \, S_{mi} S_{mj}$ is minimum.
\State Return the sign structure, $\{s_i\}$ for $s_i = S_{m^\star i}$.
\end{algorithmic}
\end{algorithm}

For GS \prp where the weights are generated from GS amplitudes, the \gw algorithm can recover the GS signs exactly. This is because for GS weights all the optimal vectors, $\mathbf{v}^\star_\sigma$, are collinear and their directions correspond to the GS sign structure. Hence the \gw algorithm can realize the \mct exactly, i.e., just like the \rt case the integrality gap is zero, and a hyperplane that does not contain the rank-$1$ solutions can reproduce the GS signs exactly. We prove this fact in the following section.

\begin{table*}[tb]
  \centering
  \small
  \setlength{\tabcolsep}{5.0pt}
\renewcommand{\arraystretch}{1.2}
  \begin{tabular*}{\textwidth}{@{\extracolsep{\fill}}lllccccccccccc}
    \toprule
    && &\multicolumn{11}{c}{$J_2/J_1$} \\
    \cmidrule(lr){4-14}
      Lattice &Size &Method
      & $0.0$ & $0.1$ & $0.2$ & $0.3$ & $0.4$ & $0.5$
        & $0.6$ & $0.7$ & $0.8$ & $0.9$ & $1.0$ \\
    \midrule
    \multirow{2}{*}{Chain}
      & \multirow{2}{*}{$16$} &\gw
      & 1.000 & 1.000 & 1.000 & 1.000 & 1.000 & 1.000
      & 1.000 & 1.000 & 1.000 & 1.000 & 1.000 \\
      &&\rt
      & 1.000 & 1.000 & 1.000 & 1.000 & 1.000 & 1.000
      & 1.000 & 1.000 & 1.000 & 1.000 & 1.000 \\
    \midrule
    \multirow{2}{*}{Square}
      & \multirow{2}{*}{$4\times4$} &\gw
      & 1.000 & 1.000 & 1.000 & 1.000 & 1.000 & 1.000
      & 1.000 & 1.000 & 1.000 & 1.000 & 1.000 \\
      &&\rt
      & 1.000 & 1.000 & 1.000 & 1.000 & 1.000 & 1.000
      & 1.000 & 1.000 & 1.000 & 1.000 & 1.000 \\
    \midrule
    \multirow{2}{*}{Triangular}
     & \multirow{2}{*}{$4\times4$} &\gw
     & 1.000 & 1.000 & 1.000 & 1.000 & 1.000 & 1.000
     & 1.000 & 1.000 & 1.000 & 1.000 & 1.000 \\
     &&\rt
     & 1.000 & 1.000 & 1.000 & 1.000 & 1.000 & 1.000
     & 1.000 & 1.000 & 1.000 & 1.000 & 1.000 \\
    \bottomrule
  \end{tabular*}
  \caption{Quality of the \gw algorithm implemented via BM factorization with $k=\lceil\sqrt{2|\mc V_\ell|}\rceil$ and the \rt relaxation method for GS \prp of \JJ HAF, at different coupling values ($J_2/J_1$), in the zero magnetization sector, for periodic lattices. Entries give the fraction of signs calculated by \gw or \rt matching the exact GS signs (up to three decimal places), so that unity corresponds to complete recovery of the sign structure. The values imply that every sign reconstruction was at least $99.95\%$ accurate. In all cases, we used the Adam~\cite{Adam} optimizer with a learning rate of $1$ and the number of iterations being $500$. We used $32$-bit floating-point numbers to reduce memory consumption. Additionally, for \gw algorithm $128$ random hyperplanes and a cosine decay schedule were used. Full implementation details are publicly available, see \textsf{Code Availability} at the end of the paper.}
  \label{tab:gw-vs-rank2}
\end{table*}

\subsection{Time Complexity of Max-Cut for Non-Degenerate GS}
The \gw relaxation offers a straightforward way to determine how difficult it is to reconstruct the sign structure using \mct when the weights come from GS amplitudes. The approach in this section does not rely on the specific form of the Hamiltonian. As long as we are working in a finite-dimensional vector space with a real Hamiltonian, the proof applies. 

In this proof we use the matrix representation of the HAF in sector $\ell$, $H_\ell$, with orthonormal basis $\mc V_\ell$. However, they can be replaced with any real Hamiltonian matrix represented in an orthonormal basis. For the rest of this section, we will also assume the wavefunction is normalized for simplicity. Since $H_\ell$ is real, its eigenvectors can also be taken as real. Let
\[
\ket{\Psi_\mathrm{GS}} = \sum_{\sigma \in \mc V_\ell} s_\sigma \psi_\sigma \ket{\sigma},
\qquad s_\sigma = \pm 1, \psi_\sigma \geq 0,
\]
be the non-degenerate GS of $H_\ell$ with energy $E_0$. When doing GS \prp using \gw algorithm, without any loss of generality we can say that the algorithm minimizes the following function with respect to $X$ in Eq.~\eqref{eq:gram_matrix}
\[
E_\gw = \sum_{\sigma, \tau \in \mathcal V_\ell} \psi_\sigma \psi_\tau (H_\ell)_{\sigma \tau} X_{\sigma \tau}
\quad \text{s.t.} \quad X_{\sigma \sigma} = 1, \, X \succeq 0.
\]
The proof proceeds as follows. First, we show that the minima of $E_\mathrm{GW}$ equal the GS energy, $E_0$. At the minima, the vectors $v_\sigma$ are collinear; thus, choosing a hyperplane that does not contain this line will reproduce the GS signs exactly. We then show that it is sufficient to reach within a small region of the minima to reproduce the GS signs. Minimizing \gw with respect to $X$ is convex, so we can always get arbitrarily close to the minima~\cite {Alizadeh1995}. Thus, the time complexity depends on the number of steps required to reach such a region.

We start by determining a lower bound on $E_\gw$. Define a symmetric matrix,
\[
\rho_{\sigma\tau} = \psi_\sigma \psi_\tau X_{\sigma\tau}.
\]
For $D_\psi = \mathrm{diag}(\psi_{\sigma_0}, \psi_{\sigma_1}, \ldots, \psi_{\sigma_{|\mathcal V_\ell|-1}})$, $\rho = D_\psi \, X \, D_\psi$. Since $X \succeq 0$ it implies $\rho \succeq 0$, as for any vector $\mu$, $\mu^T \rho \mu = (D_\psi \mu)^T X (D_\psi \mu) \geq 0$. Also, $\operatorname{Tr}(\rho) = 1$ as $\rho_{\sigma\sigma} = \psi_\sigma^2$. Thus, $\rho$ is a valid density matrix. Consider the orthonormal eigendecomposition of $\rho$, guaranteed by the spectral theorem.
\be\label{expansion of rho}
\rho = \sum_{\alpha=0}^{|\mc V_\ell| - 1} p_\alpha \ket{p_\alpha} \bra{p_\alpha},
\ee
where $\ket{p_\alpha}$ are the eigenfunctions of $\rho$ with eigenvalues $p_\alpha \geq 0$ as $\rho \succeq 0$ and $\sum_\alpha p_\alpha = \operatorname{Tr}(\rho) = 1$. Then,
\be\label{GW energy in terms of density matrix}
E_\gw = \operatorname{Tr}(H_\ell \, \rho) = \sum_\alpha p_\alpha \braket{p_\alpha | H_\ell | p_\alpha}
\ee
But the energy of any state in the vector space spanned by $\mc V_\ell$ is bounded below by the GS energy of $H_\ell$. Thus,
\[
E_\gw \geq E_0,
\]
with equality achieved only when $\ket{p_\alpha}$ are GS for all $p_\alpha \neq 0$. When the GS is non-degenerate, there is only one choice of the density matrix for the equality, and that is a pure state constructed from the GS,
\[
\rho = \ket{\Psi_\mathrm{GS}} \bra{\Psi_\mathrm{GS}}.
\]
Comparing the non-zero matrix elements, we get,
\[
X_{\sigma \tau} = \mathbf v_\sigma \cdot \mathbf v_\tau = s_\sigma s_\tau = \pm 1, \quad \forall \,\psi_\sigma \psi_\tau \neq 0
\]
If any $\psi_\sigma = 0$ then it will lead to a corresponding column and row of $\rho$ being zero. The signs for these zero amplitudes can be chosen arbitrarily without affecting the energy, hence their values do not matter. The form of $X$ above is a valid solution as it satisfies $X_{\sigma\sigma}=1$ and $X \succeq 0$. The latter follows from $X = \mathbf s \, \mathbf s^T$, where $\mathbf s$ is the column vector of GS signs. The equation above also implies that the optimal values of $\mathbf v_\sigma$ for vertices with non-zero amplitudes are collinear. In other words, $s_\sigma \mathbf v_\sigma$ points in the same direction for all $\psi_\sigma \neq 0$. Then choosing a hyperplane perpendicular to this line will reproduce the exact GS sign structure.

In reality, SDP solvers are not guaranteed to reach the minima exactly, but they are guaranteed to reach within a certain bound in finite steps. Let us assume that we are within a certain distance of the ground state energy, quantified in the error measure
\[
\epsilon = E_\gw - E_0 \geq 0.
\]
In this case, $\mathbf v_\sigma$ are not guaranteed to be collinear, and neither is $\rho$ guaranteed to be a pure state. We want $\epsilon$ small enough to still recover the GS signs. Since the hyperplane is no longer unique, we choose it to be perpendicular to a vector with nonzero amplitude, say $\mathbf v_{\sigma_r}$. Then, the condition for reproducing the GS signs is
\be\label{exact sign from gw}
s_\sigma \mathbf v_\sigma \cdot s_{\sigma_r} \mathbf v_{\sigma_r} > 0, \quad \forall \psi_\sigma \neq 0.
\ee
Consider the probability-averaged direction defined as
\[
\mathbf u = \sum_{\sigma \in \mathcal V_\ell} \psi_\sigma^2 s_\sigma \mathbf v_\sigma.
\]
We will use this direction as a reference. The GS signs $s_\sigma$ are required to construct this vector, which we do not know yet. However, we do not need to calculate the signs as the upper bound on $\epsilon$ will not depend on them, we only require the existence of the GS signs. The magnitude of $\mathbf u$ is connected with the energy, as we will show now. By the spectral theorem, let $\ket{E_i}$ be the orthonormal eigenvectors of $H$ with eigenvalues $E_i$ such that $E_{i+1} \geq E_i$. Then, using $\ket{p_\alpha} = \sum_i \braket{E_i | p_\alpha} \ket{E_i}$ in \eqref{GW energy in terms of density matrix},
\be\label{eq: E_gw in terms of E_i}
E_\gw = \sum_{i, \alpha} p_\alpha \left\lvert\braket{E_i | p_\alpha}\right\rvert^2 E_i.
\ee
Since $E_0 < E_1 \leq \ldots \leq E_{|\mathcal V_\ell| - 1}$ and $\sum_{i \geq 1} \left\lvert \braket{E_i | p_\alpha} \right\rvert^2 = 1 - \left\lvert \braket{E_0 | p_\alpha} \right\rvert^2$, we get the following inequality from Eq.~\eqref{eq: E_gw in terms of E_i} by replacing $E_{i>1}$ with $E_1$, as follows:
\be\label{upper bound on epsilon 1}
\epsilon \geq \left( 1 - \sum_\alpha p_\alpha \left\lvert \braket{E_0 | p_\alpha} \right\rvert^2 \right) (E_1 - E_0),
\ee
and the square of the magnitude of $\mathbf u$ is given by,
\begin{align*}
|\mathbf u|^2 &= \sum_{\sigma, \tau \in \mathcal V_\ell} \psi_\sigma^2 \psi_\tau^2 s_\sigma s_\tau \mathbf v_\sigma \cdot \mathbf v_\tau, \nonumber\\
&= \braket{E_0 | \rho | E_0}
= \sum_\alpha p_\alpha \left\lvert \braket{E_0 | p_\alpha} \right\rvert^2.
\end{align*}
Putting this value in \eqref{upper bound on epsilon 1} we get,
\be\label{upper bound |u|}
|\mathbf u | \geq \sqrt{1 - \frac{\epsilon}{E_1 - E_0}} \geq 1 - \frac{\epsilon}{E_1 - E_0}.
\ee
Here we assume $\epsilon < E_1 - E_0$ which also implies that $\mathbf u \neq 0$. Now, let us decompose $s_\sigma \mathbf v_\sigma$ into parallel and perpendicular parts:
\be\label{eq: parallel perp decomposition}
s_\sigma \mathbf v_\sigma = \cos\theta_\sigma \, \hat{\mathbf u} + \sin\theta_\sigma \, \hat{\mathbf n}_\sigma,
\ee
where $\hat{\mathbf u} = {\mathbf u}/{|\mathbf u|}$ and $\hat{\mathbf n}_\sigma$ is a unit vector perpendicular to $\hat{\mathbf u}$. The angle $s_\sigma \mathbf v_\sigma$ makes with $\mathbf u$ is $\theta_\sigma$. From the inequality $\hat{\mathbf n}_\sigma \cdot \hat{\mathbf n}_{\sigma_r} \geq -1$ and Eq.~\eqref{eq: parallel perp decomposition} it follows that,
\[
s_\sigma \mathbf v_\sigma \cdot s_{\sigma_r} \mathbf v_{\sigma_r} \geq \cos(\theta_\sigma + \theta_{\sigma_r}).
\]
Comparing this with \eqref{exact sign from gw} we conclude that for reconstructing the GS signs exactly, it is sufficient to satisfy
\be\label{bound on angle spread}
0 \leq \theta_\sigma < \pi / 4 \quad \forall \psi_\sigma \neq 0.
\ee
The spread in the angles $\theta_\sigma$ is directly determined by $\epsilon$, the smaller the $\epsilon$ the smaller the spread. Our task now is to determine the upper bound on $\epsilon$ such that \eqref{bound on angle spread} is maintained. Determining $|\mathbf u|$ in terms of the angles, we get the following relation
\[
|\mathbf u | = \hat{\mathbf u} \cdot \mathbf u = \sum_{\sigma \in \mathcal V_\ell} \psi_\sigma^2 \cos\theta_\sigma.
\]
Using this in Eq.~\eqref{upper bound |u|}
\[
1 - |\mathbf u | = \sum_{\sigma \in \mathcal V_\ell} \psi_\sigma^2 (1 - \cos\theta_\sigma) \leq \frac{\epsilon}{E_1 - E_0}.
\]
Since $\psi_\sigma^2 (1 - \cos\theta_\sigma)$ is always non-negative, the inequality implies that each term inside the sum cannot exceed the upper bound.
\[
\psi_\sigma^2 (1 - \cos\theta_\sigma) \leq \frac{\epsilon}{E_1 - E_0}, \quad \forall \sigma \in \mathcal V_\ell.
\]
If $\psi_\sigma=0$ then the inequality is automatically satisfied. For vertices with non-zero amplitudes,
\[
\cos\theta_\sigma \geq 1 - \frac{\epsilon}{\psi^2_\mathrm{min} (E_1 - E_0)}, \quad \forall \psi_\sigma \neq 0,
\]
where $\psi_\mathrm{min}$ is the minimum nonzero amplitude. Combining this result with \eqref{bound on angle spread} we find that the following upper bound on $\epsilon$ is enough to guarantee that GS signs will be reproduced exactly through hyperplane rounding, when the plane is perpendicular to any $\mathbf v_{\sigma}$ for $\psi_\sigma \neq 0$.
\be\label{eq: epsilon upper bound}
\epsilon < \left( 1 - \cos\frac{\pi}{4} \right) \psi_\mathrm{min}^2 (E_1 - E_0).
\ee
This relation is consistent with our previous assumption, $\epsilon < E_1 - E_0$. Now that we know $\epsilon$, reaching this accuracy can be done in polynomial time in $|\mc V_\ell|$. Theorem 3.7 of Ref.~\cite{Alizadeh1995} shows that solving for $X$ such that $E_\gw - E_0 < \epsilon$ can be done in $\mathcal O(\sqrt{|\mc V_\ell|} |\ln \epsilon|)$ steps using interior point method. Each step in interior point requires solving an $|\mc V_\ell|\times |\mc V_\ell|$ linear system which costs $\mathcal O(|\mc V_\ell|^3)$ operations~\cite{goemans1995improved}, making the total time complexity $\mathcal O \left( |\mc V_\ell|^{3.5} |\ln [\psi_\mathrm{min}^2 (E_1 - E_0)]| \right)$. Further estimates can be made by making assumptions about the scaling of the spectral gap $E_1 - E_0$ and $\psi_\mathrm{min}$ with the number of spins. In particular, we expect logarithmic corrections in $|\mc V_\ell|$ to the $|\mc V_\ell|^{3.5}$ scaling when $E_1 - E_0$ and $\psi_\mathrm{min}$ decay exponentially with number of spins. The interior point method is only used as a reference in this work, to show that \gw algorithm can indeed exactly recover the GS signs in time polynomial in HG size. Table~\ref{tab:gw-vs-rank2} uses the BM factorization due to it being much easier to implement and also due to its efficiency.

We emphasize that this solvability scaling only holds when the weights are generated from GS amplitudes and the scaling is polynomial in $|\mc V_\ell|$, the HG size, which can be exponential in the number of spins. A lower time complexity may be achieved by exploiting the structure of token graphs, as done in Ref.~\cite{ruy_inprep} for unweighted \mct. Nevertheless it is not surprising that GS \prp is exactly solvable in polynomial time in $|\mc V_\ell|$, since the GS itself can be obtained through exact diagonalization using the Lanczos algorithm in polynomial time~\cite{lanczos_complexity}. The expression for time complexity states that if the spectral gap $E_1 - E_0$ is low or the minimum non-zero amplitude is low then we will need more iterations to solve the signs. A lower spectral gap means it is harder to distinguish the GS from the first excited state, and changing the signs of low-amplitude states creates very little change in energy; thus, more iterations are needed in these cases.

\textcite{Westerhout:2022lil} used simulated annealing and a greedy algorithm to reconstruct GS signs from GS amplitudes. They found that these methods can reconstruct GS signs extremely accurately on the entire HG and on subgraphs of HG, and the sign reconstruction performs well under perturbation of the GS amplitudes. Moreover, edges with higher-valued weights satisfied PEC more often than edges with lower-valued weights. Based on these observations, they described the resulting sign optimization landscape as \textit{non-glassy} and \textit{very mildly frustrated}. 

Our results provide a rigorous counterpart to these numerical observations. Solving a general classical Ising model is \np-hard as it is equivalent to solving a general weighted \mct problem. However, we have shown that GS \prp, whose weights are generated from GS amplitudes, is exactly solvable in polynomial time in the number of vertices. The structure imposed on the weights by the GS amplitudes is enough to make the problem tractable. This suggests that the accuracy of simulated annealing and the greedy algorithm when working on the entire HG can also be attributed to the fact that the weights are generated from GS amplitudes and we conjecture that this accuracy should be reproducible on any system described by a real Hamiltonian in a finite-dimensional Hilbert space and with a non-degenerate GS. As for frustration, we have used this term as a topological property of the HG inherited from the physical graph, so our analysis has focused extensively on frustration in unweighted HG. Nevertheless, the \mct picture shows why edges with higher weights should satisfy PEC more often, since maximizing the cut value favors cutting high-weight edges. Two questions lie beyond the scope of this paper and are left for future work: defining an observable to measure the frustration of weighted HG generated from GS amplitudes and how GS \prp behaves on subgraphs of HG, or under perturbation of GS amplitudes.

\section{Sign Learning in VMC}
\label{sec: sign learning in vmc}

In previous sections, we studied the \prp where the weights of the phase-dependent interference terms were determined by fixed amplitudes. We now study the energy minimization of $H_\ell$ or equivalently maximization of the Rayleigh quotient of $L_{\ell}$, the spectral relaxation of \mct on the HG; where amplitudes are allowed to vary. VMC performs this optimization within a chosen ansatz that allows continuous probability amplitudes. But since the HAF GS is real up to a global phase, its sign structure must be well represented by this ansatz. However it has been found that NQS struggle at learning the signs, a binary classification problem, through continuous phase optimization~\cite{Westerhout:2022lil, szabo, Bukov}. As such, NQS are often paired with a sign-prior, i.e. associating a pre-defined sign $s_\sigma$ to each basis $\ket\sigma$, where the prior is a good approximation of the GS signs. Using a particular type of ansatz as an example we now show how continuous optimization of both amplitude and phase leads to a rugged optimization landscape through the inheritance of non-convex features of the \rt relaxation.

\subsection{Optimization Landscape of Complex-valued Ansatz}\label{sec:pr_vmc}

We follow the standard convention where a variational ansatz parameterizes the logarithm of the probability amplitude instead of the probability amplitude itself~\cite[Appendix A1]{Kochkov2021}. In this parameterization, the real part of a complex-valued ansatz output gives the logarithm of the wavefunction magnitude, which we call the amplitude, while the imaginary part gives its phase. This ansatz relaxes the signs to continuous phases on a unit circle, which is necessary as discrete signs are not suitable for gradient based methods. When a sign-prior is used, the ansatz only needs to determine the deviation of GS signs from the prior. This sign-prior can be incorporated either into the ansatz as an imaginary number representing the phase or equivalently into the Hamiltonian through a diagonal unitary transformation, which we discuss in details in Sec.~\ref{sec:sign_structure_MC}. We use the former method for convenience.

The amplitude and phase represented by a complex-valued NQS may depend on shared or separate variational parameters~\cite{lange2024architectures}. We choose the latter as VMC with this type of ansatz can be shown to have the most direct connection to \prp. We denote the amplitude and phase parameters by the tuples $\theta$ and $\alpha$, respectively. We also make the deliberate choice of representing the amplitude as $\psi(\theta)$ instead of $e^{\psi(\theta)}$, for notational convenience. The variational ansatz then takes the form,
\[
\ket{\Psi_{\theta\alpha}} = \sum_{\sigma \in \mc V_\ell} \psi_\sigma(\theta) \, e^{i \phi_\sigma(\alpha)} \ket \sigma, \quad \psi_\sigma > 0, \phi_\sigma \in \mathbb R.
\]
We work with the untransformed Hamiltonian in $\ell = \lfloor N/2 \rfloor$ sector and the sign-prior, if used, is accounted for by adding the phase associated with the sign-prior to the imaginary part of the ansatz output. Let
\[
(w_\ell)_{\sigma\tau} \, (\theta) = 2 \, H_{\sigma\tau} \, \frac{\psi_\sigma(\theta) \psi_\tau(\theta)}{\braket{\Psi_{\theta\alpha}|\Psi_{\theta\alpha}}} , \quad \forall \{\sigma,\tau\} \in \mc E_\ell,
\]
be the amplitude-dependent weights, which depend only on $\theta$. The classical and quantum parts of the variational energy then become,
\begin{align*}
E_c(\theta) &= \sum_{\sigma \in \mc V_\ell} \frac{\psi_\sigma(\theta)^2 H_{\sigma\sigma}}{\braket{\Psi_{\theta\alpha} | \Psi_{\theta\alpha}}}, \\
E_q(\theta, \alpha) &=
\sum_{\{\sigma,\tau\} \in \mc E_\ell} (w_\ell)_{\sigma\tau}(\theta) \cos\big[\phi_\sigma(\alpha) - \phi_\tau(\alpha)\big].
\end{align*}
This expression for $E_q$ differs from Eq.~\eqref{eq:5} in two ways: the weights are no longer constant, and the phases are also parameterized. The variational energy is the sum of the two parts, $E(\theta, \alpha) = E_c(\theta) + E_q(\theta, \alpha)$. In VMC, quantities such as the variational energy and its gradient are estimated by averaging local estimators corresponding to the quantities over configurations sampled from the Born distribution~\cite{sorella}. In our case, we only need that VMC provides a stochastic estimate of the variational energy and its gradient. Therefore, we can infer what VMC does by analyzing $E_c$, $E_q$, and their gradients. Interestingly, the Markov Chain Monte Carlo sampling can be viewed as a random walk on HG with Metropolis–Hastings transition probabilities because magnetization-preserving sampling generates samples using spin exchanges or HFs along the edges of the physical lattice.

When stochastic gradient descent is used, the parameters are updated at every iteration in the opposite direction of the energy gradient. Other gradient-based methods such as stochastic reconfiguration~\cite{sorella} and Adam~\cite{Adam} modify this scheme by transforming the gradient before using it to update the parameters. Analysis of stochastic gradient descent therefore has implications for other methods as well. Now, the phase parameters $\alpha$ are updated only through derivatives of variational energy with respect to $\alpha$ itself, i.e. $\nabla_\alpha E$, and likewise for $\theta$. Since $E_c$ is independent of $\alpha$, $\nabla_\alpha E = \nabla_\alpha E_q$. So the parameter updates become,
\[
\theta \gets \theta - \zeta \, \nabla_\theta E(\theta, \alpha), \qquad
\alpha \gets \alpha - \zeta \, \nabla_\alpha E_q(\theta, \alpha).
\]
Here, $\zeta$ is the learning rate. Expanding $\nabla_\alpha E_q$ gives,
\be\label{eq: grad_beta E_q}
\nabla_{\alpha} E_q = - \sum_{\{\sigma,\tau\}\in\mc E_\ell} (w_\ell)_{\sigma\tau} \sin\left(\phi_\sigma - \phi_\tau\right) \nabla_{\alpha} \left( \phi_\sigma - \phi_\tau \right).
\ee
This is equivalent to Eq.~\eqref{eq: dE_q/dphi} with the phases being parameterized by an ansatz. Thus, at each iteration the phase parameters are updated as if it is solving a \prp on the HG with weights defined by the current $\theta$. That is to say, when amplitude and phase use separate parameters, VMC can be considered as an evolving \rt relaxed \mct problem: The amplitudes and weights evolve during training, and at each step the phases $\phi_\sigma(\alpha)$ are updated as in the \rt relaxation method for \mct with weights $(w_\ell)_{\sigma\tau} \, (\theta)$ on the HG. Since $\alpha$ are updated using $\nabla_\alpha E$, we need to analyze the optimization landscape of $E(\theta, \alpha)$ at constant $\theta$ to determine the phase learning behavior of the ansatz. Note that this interpretation is only valid when separate parameters are used for phase and amplitude.

The phase ansatz is a map from the parameter space of $\alpha$ to the space of phases: the ansatz takes the parameters $\alpha$ and a HG vertex $\sigma$ as input and outputs the phase $\phi_\sigma(\alpha)$. In the \rt method, the phases $\phi_\sigma$ are themselves variational parameters, whereas in VMC the phases are parameterized by $\alpha$. We now show how continuous parameterization of the phase makes the ansatz inherit the non-convex features of \rt relaxation method, such as the spurious minima and saddle points. We only provide a sketch of the proof; for a rigorous approach, see Proposition 3.6 and Lemma 3.11 of Ref.~\cite{Levin2025}.

Let $\Phi$ be a point in the space of phases with coordinates $\Phi_\sigma$ that is a spurious minimum of the \rt objective, Eq.~\eqref{eq: rank-2 objective}, with weight matrix $w_\ell(\theta)$. Let $\Phi$ also be reachable by the ansatz, i.e., let there exist a point $\alpha_\Phi$ in parameter space such that $\phi_\sigma(\alpha_\Phi) = \Phi_\sigma$ for all $\sigma \in \mc V_\ell$. Since the ansatz is a continuous map, a small change of parameters, $\alpha = \alpha_\Phi + d\alpha$, leads to small change of the phases $\phi_\sigma(\alpha_\Phi+d\alpha) = \Phi_\sigma + d\Phi$. Since $\Phi$ is a local minimum, no point in a sufficiently small neighborhood of $\Phi$ has a lower $E_q$. For small enough $d\alpha$, $\Phi + d\Phi$ lies in this neighborhood, so $E(\theta, \alpha_\Phi + d\alpha) \geq E(\theta, \alpha_\Phi)$. This is true for small deviations in any direction in the parameter space. Hence, $\alpha_\Phi$ is a local minimum of the energy with respect to $\alpha$, and if the ansatz can represent energies lower than $E(\theta, \alpha_\Phi)$, it is a spurious minimum. Each minimum is part of a line in the space of phases as shifting every phase by a constant does not affect the energy functional. These lines then become curves in parameter space of $\alpha$ for continuous mapping.

We now move to saddle points. Continuity alone does not imply that the ansatz is capable of representing two directions in parameter space corresponding to an increase and a decrease in energy. If the ansatz is only continuous but cannot represent both directions, then it may only represent the direction of increase or only the direction of decrease. In that case, the saddle point becomes a spurious minimum or a local maximum. Let $\Phi$ be a saddle point of the \rt objective, Eq.~\eqref{eq: rank-2 objective}. For sufficiently small deviation $d\Phi$ there exists $d\alpha$ such that $\phi_\sigma(\alpha_\Phi + d\alpha) = \Phi_\sigma + d\Phi_\sigma$. At a saddle point, there exist at least two directions $d\Phi_1$ and $d\Phi_2$ such that energy decreases along the first and increases along the second. Under our local expressivity assumption, these two directions have corresponding counterparts $d\alpha_{1}$ and $d\alpha_{2}$ in parameter space, respectively. Therefore, $E(\theta, \alpha_\Phi + d\alpha_1) < E(\theta, \alpha_\Phi) < E(\theta, \alpha_\Phi + d\alpha_2)$, and $(\theta, \alpha_\Phi)$ is a saddle point of energy with respect to $\alpha$ itself.

As discussed in Sec.~\ref{sec: rank-2 relaxation}, every $\{0, \pi\}$ phase assignment, $\phi_\sigma \in \{0, \pi\}$ is a stationary point of the \rt objective, Eq.~\eqref{eq: rank-2 objective} and if the $\{0, \pi\}$ assignment does not minimize or maximize the energy globally then it is a saddle point. There are at most $2^{n-1}$ lines that correspond to such saddle points, where $n = |\mc V_\ell|$. By Eq.~\eqref{eq: dE_q/dphi} these $\{0, \pi\}$ assignments are also stationary points of the phase ansatz. For an expressive ansatz, these lines of saddle points correspond to curves in the parameter space of $\alpha$. Since this stationarity does not depend on $\theta$, each line becomes a manifold of dimension $|\theta|+1$ in the combined parameter space of $\theta$ and $\alpha$. These manifolds differ from the stationary points that represent energy eigenstates, which form lower-dimensional manifolds in parameter space, as only the global phase and the normalization can be chosen freely.

The image of the phase ansatz in the space of phases is at most $|\alpha|$ dimensional and in practice $|\alpha| \ll |\mc V_\ell|$. Therefore the local expressivity only works inside the much lower dimensional image, i.e. points in space of phases exactly representable by the ansatz. The phase ansatz then misses most parts of the lines of $\{0, \pi\}$ assignments and only exactly representable minima from \rt objective are inherited. Moreover, the ansatz only sees the Hessian of $E_q$ with respect to $\alpha$, which is not the full picture. The ansatz can at most represent $|\alpha|$ independent directions at stationary points which is a small fraction of the $|\mc V_\ell|$ dimensional space. So saddle points whose descent directions are not representable become spurious minima. Let $\nabla_\alpha E_q = J^\top \nabla_\phi E_q$, where $J = \partial \phi / \partial \alpha$ is the Jacobian and has rank at most $|\alpha|$. So, a point in parameter space, $\alpha$, is also stationary whenever $\nabla_\phi E_q$ is orthogonal to the image of $J$, which creates stationary points with no counterpart in the \rt objective, while nearly reachable stationary lines produce plateaus in parameter space. Suffice to say, the optimization landscape remains highly non-convex, consistent with the rugged landscape reported by \textcite{Bukov}, who also used separate parameters for amplitude and phase. Our \prp formalism has therefore allowed us to connect the rugged landscape of VMC with the non-convex features of \rt relaxation.

This formalism also allows for a possible explanation of why sign learning is observed to be easier on bipartite lattices~\cite{Westerhout:2019hvt}. When the physical graph is bipartite the HG is also bipartite and the \mct on HG is therefore independent of the amplitude values, solely determined by the MSR. At each iteration, no matter the value of $\theta$, the MSR globally minimizes $E_q(\theta,\alpha)$. The optimal phases therefore do not move as $\theta$ changes. Assuming each update direction on average points towards the current global minima of $E_q(\theta, \alpha)$, for a bipartite lattice $\alpha$ gets pushed towards the direction of MSR at every iteration. So the sign structure should be learned faster than the non-bipartite case.

\subsection{Interpretation of VMC with Real-valued Ansatz}

We now choose a real-valued ansatz which by the convention used, cannot represent the phases and thus the ansatz must be combined with a sign prior. In this case there is no direct relationship with a continuous relaxation of \mct, therefore we will not be able to discuss the optimization landscape as before. However, we can at least show how VMC in this case can be viewed through the lens of HG formulation. Let $\theta$ denote the variational parameters, then the variational wavefunction in this case effectively has the following form.
\[
\ket{\Psi_{\theta}} =  \sum_{\sigma \in \mc V_\ell} s_\sigma \psi_\sigma(\theta) \ket{\sigma},
\quad \psi_\sigma(\theta) > 0
\]
The sign-prior enters the equation through $s_\sigma$ and $\ell = \lfloor N / 2 \rfloor$. The definition of parameter dependent weight matrix $w_\ell(\theta)$ has the same form as before and so does $E_c$. The quantum part of variational energy, however, is different as there are no interference terms.
\[
E_q(\theta) = 
\sum_{\{\sigma,\tau\} \in \mc E_\ell} (w_\ell)_{\sigma\tau}(\theta) \, s_\sigma s_\tau 
\]
Here, $s_\sigma s_\tau$ comes from the sign-prior. Now, a sign-prior defines a sign for each HG vertex and therefore defines a cut in HG. Thus, when a real ansatz is used with a sign prior, VMC optimization works with a pre-defined cut. This realization allows us to see how energy optimization using a real ansatz is constrained by the sign-prior from the HG perspective. The sign prior pre-defines the cut of HG and then VMC has to find the optimal parameters such that the energy is minimized. To get a good approximation of the GS energy the pre-defined cut has to match the \mct formed by GS signs as closely as possible with more priority given to cutting high-weight edges. Therefore getting the signs of states with larger amplitudes right is important. However, for HAF, ignoring special coupling values, correct expressions for the sign-priors are only known when the graph is bipartite. For these cases, PEC is satisfied globally and the \mct is independent of the weights, fixed by the MSR; so no matter the value of the amplitudes, VMC always works with a \mct.

\section{Stoquasticity and \mct}\label{sec:sign_structure_MC}
In the previous sections, we have studied the sign structure through the phases of the variational wavefunction. We now approach the sign structure from the Hamiltonian side by examining its off-diagonal matrix elements in the computational basis.  In the computational basis $\{\ket{\sigma}\}$, a Hamiltonian $\hat H$ is stoquastic~\cite{BDOT2008} if
\[
\braket{\sigma | \hat H | \tau} \leq 0, \qquad \text{for every}\, \sigma\neq \tau.
\]
We show that a diagonal sign transformation can make $H_\ell$ stoquastic when the \mct deficit on coupling-weighted HG vanishes.

The token graph structure of the HG shows the basis dependence for representing the Hamiltonian. In the computational basis, each state specifies the positions of $\ell$ up spins, which we take as tokens. An HF moves a single token along a physical bond to a site occupied by a down spin. These spin flips establish the isomorphism $\Gamma_\ell(G)\cong F_\ell(G)$. Each HF can therefore be viewed as a token hop between neighboring vertices of the HG, with hopping amplitude $J_r/2$ for a physical bond of range $r$. The diagonal terms give the energy of the spin configuration at each vertex. In sector $\ell$, the HAF Hamiltonian in Eq.~\eqref{eq:Ham} takes the form
\be\label{eq:H_HG}
\begin{split}
H_\ell &= \sum_r \frac{J_r}{2} \sum_{\{\sigma,\tau\}\in\mathcal E_{\ell, r}}\bigg[\ket{\sigma}\bra{\tau}+\ket{\tau}\bra{\sigma}\bigg] \\
&\quad +\sum_{\sigma\in\mathcal V_\ell}\left[E_F - \sum_r\frac{J_r}{2}n_r^{\rm ap}(\sigma)\right]\ket{\sigma}\bra{\sigma}.
\end{split}
\ee
Equation~\eqref{eq:H_HG} gives a tight-binding representation of the HAF
Hamiltonian on the HG. Here, the hopping term follows from the off-diagonal part of each bond operator $J_r\,\hat{\mathbf S}_i\cdot\hat{\mathbf S}_j$ in $G$, which exchanges antiparallel spins on the bond, and $n_r^{\mathrm{ap}}(\sigma)$ is defined in Eq.~\eqref{eq:anti-parallel-config}. In this representation, $H_\ell$ is stoquastic if all hopping amplitudes are real and nonpositive, i.e., if every hop has the standard tight-binding form $-t$ with $t\ge 0$. Since every hopping in Eq.~\eqref{eq:H_HG} equals $+J_r/2>0$, $H_\ell$ is not stoquastic in the computational basis. Writing $+J_r/2=-(J_r/2)\,e^{i\pi}$ shows that, relative to the standard form, every HG edge carries a Peierls phase $\pi$.

We now ask whether a change of basis can remove these Peierls phases.
A unitary transformation $U_\ell$ that transforms $H_\ell \to U_\ell H_\ell U_\ell^\dagger$ leaves the spectrum, and hence the GS energy, unchanged, but it can change the on-site energies and hoppings on the HG.  For a general $U_\ell$, however, the nonzero off-diagonal elements of $U_\ell H_\ell U_\ell^\dagger$ need not correspond to single-token moves along bonds of $G$, and the resulting graph need not be isomorphic to $F_\ell(G)$.

We first consider a diagonal sign transformation, which preserves the HG connectivity, hopping magnitudes, and onsite energies, and changes only the hopping phases. In sector $\ell$, it has the form
\be\label{eq:diagonal_sign_unitary}
U_\ell = \sum_{\sigma \in \mc V_\ell} s_\sigma \ket{\sigma} \bra{\sigma}, \quad s_\sigma = e^{i\phi_\sigma} \in \{\pm 1\}.
\ee
Since $s_\sigma=\pm1$, this unitary satisfies $U_\ell^\dagger=U_\ell=U_\ell^{-1}$. It assigns the sign $s_\sigma$ to each basis state: $U_\ell|\sigma\rangle=s_\sigma|\sigma\rangle$. Now applying $U_\ell$ to $H_\ell$ gives
\begin{align}\label{eq:gauge_transformed_HG_Hamiltonian}
H'_\ell & = U_\ell H_\ell U_\ell \nonumber \\
& = \sum_r \frac{J_r}{2} \sum_{\{\sigma,\tau\}\in\mc E_{\ell, r}} s_\sigma s_\tau  \bigg[\ket{\sigma}\bra{\tau}+\ket{\tau}\bra{\sigma}\bigg] \nonumber \\
& + \sum_{\sigma\in\mathcal V_\ell}\left[E_F - \sum_r\frac{J_r}{2}n_r^{\rm ap}(\sigma)\right]\ket{\sigma}\bra{\sigma}.
\end{align}
Since $J_r>0$, the transformed hopping amplitude
$(J_r/2)s_\sigma s_\tau$ has the sign of $s_\sigma s_\tau$.
Thus $H_\ell'$ is stoquastic if and only if
\be\label{eq:pec_sign}
s_\sigma s_\tau=-1 \qquad \text{for every }\{\sigma,\tau\}\in\mathcal{E}_\ell.
\ee
In the language of Peierls phases, $s_\sigma s_\tau=e^{i(\phi_\sigma-\phi_\tau)}$, so $U_\ell$ adds the phase difference $\phi_\sigma-\phi_\tau$ (equal to $\pi$ on PEC edges and $0$ on ZEC edges) to the phase $\pi$ on each edge, and Eq.~\eqref{eq:pec_sign} requires this difference to cancel the $\pi$ on every edge. This edgewise cancellation is equivalent to the cycle condition of Ref.~\cite{Hen2021}, specialized to the HAF, where every hop carries the same phase. Following Theorem~\ref{thm:bipartite_inheritance}, this is possible only
when the physical lattice is bipartite, and the diagonal unitary can then be chosen explicitly as the Lieb--Mattis operator~\cite{LM1,LM2}
\[
\eta = (-1)^{N_A^\uparrow} = (-1)^{|A|}\prod_{i\in V_A}\sigma_i^z.
\]
Here the bipartite physical graph is $G=(V_A \sqcup V_B, E)$ with disjoint sublattices $V_A$ and $V_B$, and $\hat{N}^{\uparrow}_A=\sum_{i\in V_A}\tfrac{1}{2}(I+\hat{\sigma}^z_i)$ counts the number of up-spins on $V_A$~\eqref{eq:N_A}. Restricted to sector $\ell$, $\hat\eta$ is the unitary $U_\ell$ in Eq.~\eqref{eq:diagonal_sign_unitary} that makes the bipartite-graph Hamiltonian stoquastic.
\be\label{eq:LM_mct}
\eta\ket\sigma = (-1)^{N_A^\uparrow(\sigma)}\ket\sigma =
\begin{cases}
+\ket{\sigma}, & N_A^\uparrow(\sigma)\ \text{even},\\
-\ket{\sigma}, & N_A^\uparrow(\sigma)\ \text{odd}.
\end{cases}
\ee
Comparing Eq.~\eqref{eq:LM_mct} with $U_\ell|\sigma\rangle=s_\sigma|\sigma\rangle$ gives
\[
s_\sigma=(-1)^{N_A^\uparrow(\sigma)},\qquad \phi_\sigma=\pi N_A^\uparrow(\sigma),
\]
which is the Marshall sign in Eq.~\eqref{eq:msr_sign_one} and the PEC Eq.~\eqref{eq:7}. On a bipartite lattice, every HF changes $N_A^\uparrow$ by one. The LM signs therefore differ across every HG edge, giving a cut that contains \textbf{all} edges and is a \mct for any nonnegative edge weights (Fig.~\ref{fig:lattice-and-cg}). The same sign assignment makes every hopping amplitude negative, so the transformed Hamiltonian $H_\ell'=\eta H_\ell\eta$ is stoquastic.

On a non-bipartite lattice, however, no binary sign assignment can
cut every HG edge. The cut edges satisfy the PEC condition, while
the uncut edges satisfy the ZEC given by
\begin{align*}
\mc{E}_{\mathrm{PEC}}
&= \bigl\{\{\sigma,\tau\}\in\mc{E}_\ell \,\big|\, s_\sigma s_\tau=-1\bigr\},\\ 
\mc{E}_{\mathrm{ZEC}} &= \bigl\{\{\sigma,\tau\}\in\mc{E}_\ell \,\big|\, s_\sigma s_\tau=+1\bigr\}.
\end{align*}
Consequently the \mct value for coupling-weighted adjacency decreases from the upper bound $(W_\ell)_\mathrm{tot} = \sum_{\{\sigma,\tau\} \in \mc E_\ell} (W_\ell)_{\sigma\tau}$. The off-diagonal elements that remain positive represent ZEC edges in HG. A measure of these residual positive off-diagonal weights is
\[
R(U_\ell) = \sum_{\{\sigma ,\tau\} \in \mc E_\mathrm{ZEC}} (W_\ell)_{\sigma\tau} = (W_\ell)_\mathrm{tot} - \mathrm{Cut}_{W_\ell}(s).
\]
We call this the \textit{cut deficiency}. Then,
\[
\min R(U_\ell) = (W_\ell)_\mathrm{tot} - \mct(W_\ell)
\]
is the \mct deficiency. The following theorem follows from this definition.
\begin{theorem}
A diagonal real unitary transformation $U_\ell$ that can make $H_\ell' = U_\ell H_\ell U_\ell$ stoquastic exists if and only if the \mct deficiency $\min R(U_\ell) = 0$.
\end{theorem}
\begin{proof}
The transformed Hamiltonian $H_\ell'$ is stoquastic when all off-diagonal elements are non-positive. Hence $s_\sigma s_\tau = -1$ on every edge and PEC is obeyed globally. Thus, no edge satisfies the ZEC conditions and the MSR gives $R(U_\ell) = 0$, which is also the minimum of $R(U_\ell)$ as the weights are non-negative. Conversely, when there exists a sign assignment $s_\sigma$ such that $R(U_\ell) = 0$ then due to all $(W_\ell)_{\sigma\tau}$ being non-negative for HAF, we get global PEC which implies $H_\ell'$ is stoquastic.
\end{proof}

By the Perron-Frobenius theorem, when a unitary transformation renders the Hamiltonian stoquastic, the GS becomes positive, i.e., all signs of the GS probability amplitudes can be chosen to be $+1$. Thereby, $R(U_\ell)=0$ implies a positive ground state. However, deviation of $R(U_\ell)$ from zero is not guaranteed to break this positivity, even though it breaks the stoquasticity. 

In $J_1$--$J_2$ models, the exact LM transformation can still survive at special coupling limits. For example, in the square-lattice $J_1$--$J_2$ HAF, $G = (V,E_1)$ at $J_2=0$ and $G = (V,E_2)$ at $J_1=0$; at both limits the physical graph is bipartite. Assuming the probability amplitudes vary continuously with $J_2/J_1$, it is reasonable to expect the sign structure at a particular $J_2/J_1$ to persist near that coupling and the signs of the states with larger amplitudes should persist longer. In terms of \mct, we can argue that the weights also change continuously, hence the \mct at a coupling value should work as a very accurate estimation of the \mct for neighboring couplings as well. Therefore near each of these limits the canonical LM parity operator, $\eta$ provides a natural sign prior, where $V_A$ is a disjoint sublattice of $(V, E_1)$ for $J_2 \ll J_1$ and disjoint sublattice of $(V, E_2)$ for $J_2 \gg J_1$~\cite{CNN1}. Due to the anticommutation of Pauli operators, the Hamiltonian is transformed as,
\be\label{eq: H' vivj}
\hat H'
= \sum_{r=1}^2 J_r \sum_{\{i, j\} \in E_r} \left[ \hat S^z_i \hat S^z_j + v_{i}v_{j} \left( \hat S^x_i \hat S^x_j + \hat S^y_i \hat S^y_j \right) \right],
\ee
where $v_{i} = -1$ whenever $i \in A$, otherwise $v_{i} = +1$.
Eq.~\eqref{eq: H' vivj} and Eq.~\eqref{eq:gauge_transformed_HG_Hamiltonian} are related by HFs.
\[
s_\sigma s_\tau = v_i v_j, \quad \forall \, \sigma \overset{ij}{\rightleftharpoons} \tau.
\]
The transformation shows that for any nonzero finite $J_2/J_1$, some off-diagonal elements remain positive, breaking stoquasticity even though we can argue perturbatively that the GS remains positive in special limits. Beyond such special cases, however, no analogous \textit{a priori} grading determined solely by the lattice structure is generally available. Nevertheless, this discussion contrasts the quantum Monte Carlo (QMC) sign problem~\cite{Sandvik_2010,pan2024sign} with the VMC sign learning problem. Solving the QMC sign problem always solves the VMC sign-learning problem if the ansatz takes the rotated basis states as input. But solving the VMC sign-learning problem only requires a transformation that maps the GS to a positive vector, which may not make the Hamiltonian stoquastic.

The diagonal unitary transformations considered here preserve the HG connectivity, and hence its token-graph structure, while changing only the phases of the off-diagonal Hamiltonian matrix elements.  From this viewpoint, searching for a stoquastic representation means constructing a new weighted HG for the unitarily transformed Hamiltonian, with real, nonpositive off-diagonal matrix elements. The resulting HG need not retain the token-graph structure. The problem is therefore to find efficiently constructible non-diagonal transformations that produce a non-token stoquastic HG at fixed system size and magnetization, while keeping the transformed Hamiltonian efficiently representable. This connects our graph perspective to the broader search for stoquastic bases studied by Marvian \textit{et al.}~\cite {Marvian_2019}. We leave these transformations and the structure of the resulting HGs for future work.

\section{Average Sign and Estrada bipartivity}\label{sec:average_sign}

In this work, we have studied the sign-learning difficulty of NQS in VMC from the HG viewpoint and used Estrada’s bipartivity index in Subsec.~\ref{subsec:estrada} to quantify the HG’s departure from bipartiteness. The even and odd closed walks underlying this index also determine the average sign in QMC. For the HAF, we derive an exact relation between the average sign and a temperature-dependent generalization of Estrada’s index that retains the Hamiltonian’s diagonal contributions.

Using the Hamiltonian--Laplacian identity in Eq.~\eqref{eq:H_laplacian_shifted}, the sector Hamiltonian is $H_\ell=E_F I-L_\ell$, where
$L_\ell=D_\ell-W_\ell$ as defined in Eq.~\eqref{eq:weighted_HG_laplacian}.
Here, $W_\ell$ is the coupling-weighted adjacency matrix and $D_\ell$ is its weighted degree matrix in Eq.~\eqref{eq:degree_matrix}. Replacing every off-diagonal element of $H_\ell$ by minus its absolute value changes $W_\ell$ to $-W_\ell$.
The stoquastic partner is therefore
\be\label{eq:Hsq_signless}
H_{\ell,\mathrm{sq}} = E_F - Q_\ell, \qquad Q_\ell = D_\ell + W_\ell,
\ee
where $Q_\ell$ is obtained by removing signs from the Laplacian. The average sign, $\braket{s}_{\ell, \beta}$, is the ratio of the partition function of HAF Hamiltonian, $\operatorname{Tr}(e^{-\beta H_\ell})$ and the partition function of its stoquastic partner, $\operatorname{Tr}(e^{-\beta H_{\ell, \mathrm{sq}}})$~\cite{pan2024sign}. The constant $E_F$
contributes the same factor $e^{-\beta E_F}$ to both partition functions and cancels in the ratio. In terms of $L_\ell$ and $Q_\ell$, the average sign $\langle s\rangle_{\ell,\beta}$ is given by
\be\label{eq:average_sign_trace}
\langle s\rangle_{\ell,\beta}
=\frac{\mathrm{Tr}\left[e^{\beta L_\ell}\right]}
      {\mathrm{Tr}\left[e^{\beta Q_\ell}\right]}
=\frac{\mathrm{Tr}\left[e^{\beta(D_\ell-W_\ell)}\right]}
      {\mathrm{Tr}\left[e^{\beta(D_\ell+W_\ell)}\right]}.
\ee
To separate the closed sequences according to their number of HG hops,
we expand the exponentials in Eq.~\eqref{eq:average_sign_trace}:
\[
\mathrm{Tr} \left[e^{\beta(D_\ell\pm W_\ell)}\right] = \sum_{n=0}^{\infty}\frac{\beta^n}{n!} \mathrm{Tr}\left[(D_\ell\pm W_\ell)^n\right].
\]
At order $n$, distributing the factors produces ordered products,
which we call operator words,
\[
\mathbb W_\ell = X_1 X_2\cdots X_n, \qquad X_j\in\{D_\ell,W_\ell\}.
\]
Let $\mathscr W_n$ denote the set of these words, and $N_W(\mathbb W_\ell)$ the number of $W_\ell$ factors in $\mathbb W_\ell$. Each word appears with a factor $(-1)^{N_W(\mathbb W_\ell)}$ in the expansion of $(D_\ell-W_\ell)^n$, while every word has a positive coefficient in $(D_\ell+W_\ell)^n$. Hence, we can write
\[
\mathrm{Tr}\, \mathbb W_\ell = \sum_{\sigma_0,\ldots,\sigma_{n-1}}
\prod_{j=1}^{n} (X_j)_{\sigma_{j-1}\sigma_j}, \qquad \sigma_n=\sigma_0.
\]
The trace therefore forces each contributing sequence to return
to its starting HG vertex. Since $D_\ell$ is the diagonal degree matrix,
a factor of $D_\ell$ leaves the configuration unchanged and contributes
the weighted degree $D_{\sigma\sigma}$. An $W_\ell$ factor moves between
adjacent configurations in HG and contributes the edge weight $(W_\ell)_{\sigma\tau}$. By Eq.~\eqref{eq:full_weighted_adj_matrix}, this weight is $J_r/2$ for $\{\sigma,\tau\}\in\mc E_{\ell,r}$. Thus, every contributing sequence in
$\mathrm{Tr}\,\mathbb W$ contains exactly $N_W(\mathbb W_\ell)$ hops.

We classify a closed sequence as even or odd according to this
hopping number. Diagonal stays do not affect its parity.
For example, $D_\ell^3$ and $D_\ell W_\ell^2$ belong to the even
sector, whereas $D_\ell W_\ell^3$ belongs to the odd sector despite
appearing at fourth order in $\beta$.
We now define the total weights of the two sectors:
\begin{align*}
E_\ell(\beta) & = \sum_{n=0}^{\infty}\frac{\beta^n}{n!}\sum_{\substack{
\mathbb W_\ell \in\mathscr W_n\\ N_W(\mathbb W_\ell)\ \mathrm{even}}} \operatorname{Tr} \mathbb W_\ell, \\
O_\ell(\beta) &= \sum_{n=0}^{\infty}\frac{\beta^n}{n!} \sum_{\substack{
\mathbb W_\ell\in\mathscr W_n\\ N_W(\mathbb W_\ell)\ \mathrm{odd}}}\operatorname{Tr} \mathbb W_\ell.
\end{align*}
All contributions to these sums are nonnegative for $\beta\geq0$.
The empty word at $n=0$ is the identity and belongs to the even
sector, giving $E_\ell(0)=|\mc V_\ell|$ and $O_\ell(0)=0$. Consequently,
\begin{align*}
    \operatorname{Tr}\left[ e^{\beta L_\ell} \right]
    &=E_\ell(\beta)-O_\ell(\beta),
    \\
    \operatorname{Tr}\left[ e^{\beta Q_\ell}\right]
    &=E_\ell(\beta)+O_\ell(\beta).
\end{align*}
This even and odd decomposition is the same one that underlies Estrada's
spectral bipartivity~\cite{Estrada}, introduced in
Sec.~\ref{subsec:estrada} and denoted there by $\beta$. Since we now use $\beta$ to denote the inverse temperature, we do not use that symbol for the Estrada index. Instead, we define a temperature-dependent
generalization that includes the diagonal stays,
\be\label{eq:generalized-bipartivity}
\mc B_\ell(\beta) = \frac{E_\ell(\beta)}{E_\ell(\beta)+O_\ell(\beta)}.
\ee
This generalized Estrada bipartivity index measures the fraction of the total absolute weight carried by
closed sequences with an even number of hops. Substituting into
Eq.~\eqref{eq:average_sign_trace}, we obtain
\be\label{eq:average-sign-bipartivity}
\langle s\rangle_{\ell,\beta} = \frac{E_\ell(\beta)-O_\ell(\beta)}
{E_\ell(\beta)+O_\ell(\beta)} = 2\mc B_\ell(\beta)-1.
\end{equation}
The relation is exact and retains the full contribution of the degree matrix $D_\ell$.

\begin{figure*}[htbp]
\centering
\includegraphics[width=1.00\textwidth]{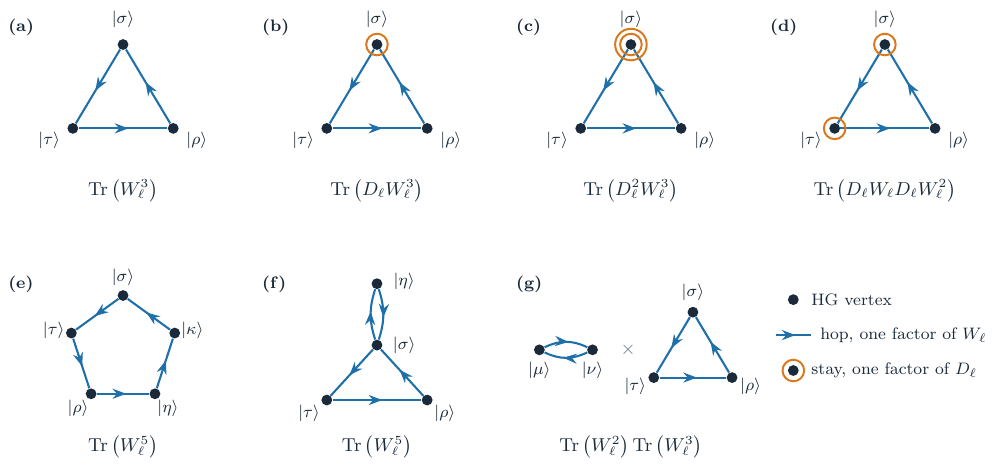}
\caption{Representative odd closed walks on the HG that contribute to the
odd-to-even walk ratio $R_\ell (\beta)$. Filled dots are HG vertices
$\ket{\sigma}$; blue arrows are hops along HG edges, each carrying one
factor of $W_\ell$, and an orange ring marks a diagonal insertion at that
vertex, carrying one factor of $D_\ell$. A ring is not an edge of the HG.
\textbf{(a)} The triangle walk $\ket{\sigma}\to\ket{\tau}\to\ket{\rho}\to\ket{\sigma}$
contributing to $\operatorname{Tr}(W_\ell^{3})$ at order $\beta^{3}$.
\textbf{(b)}--\textbf{(d)} The same triangle with one or two diagonal insertions, contributing at orders $\beta^{4}$ and $\beta^{5}$.
\textbf{(e)}--\textbf{(g)} The three five-hop structures at order $\beta^{5}$: a simple five-cycle, a triangle with one edge traversed back and forth, where $\ket{\eta}$ may also be a vertex of the triangle, and the product $\operatorname{Tr}(W_\ell^{2})\operatorname{Tr}(W_\ell^{3})$ that enters $r_{5}$ through the elimination of the even-walk series. Each trace sums over all vertex labels, and numerical prefactors are omitted.
}
\label{fig:odd_walk_HG}
\end{figure*}

For a finite HG and a finite inverse temperature $\beta\geq0$, the nonnegative walk weights imply the following.
\begin{theorem}\label{thm:odd-to-even-walk}
Let $\Gamma_\ell$ be a finite, undirected HG with nonnegative edge weights, coupling weighted adjacency matrix $W_\ell$ and degree matrix $D_\ell$. Let $E_{\ell}(\beta)$ and $O_{\ell}(\beta)$ denote the even and odd hopping weights defined above. Then, for every finite $\beta \geq 0$, the ratio
\[
R_\ell (\beta) = \frac{O_\ell(\beta)}{E_\ell(\beta)}.
\]
is well-defined and satisfies
\be\label{eq:rl_bound}
0 \leq R_\ell (\beta) < 1
\ee
\end{theorem}
\begin{proof}
Since $\Gamma_\ell$ is finite, $D_\ell$ and $W_\ell$ are finite matrices with nonnegative entries. Every contribution of $E_{\ell}(\beta)$ and $O_{\ell}(\beta)$ is therefore nonnegative for $\beta\geq 0$, and both series converge because their sum is
$E_\ell(\beta)+O_\ell(\beta)=\mathrm{Tr}\,[e^{\beta Q_\ell}] \in \mathbb{R}$. The empty operator word belongs to the even sector and contributes $\mathrm{Tr}\, I = |\mc V_\ell| >0 $. Therefore
\[
E_{\ell}(\beta) \geq |\mc V_\ell| >0, \qquad O_{\ell}(\beta) \geq 0.
\]
Moreover, writing $L_\ell = D_\ell - W_\ell$, the even–odd decomposition gives
\[
E_{\ell}(\beta) - O_{\ell}(\beta) = \mathrm{Tr}\left[e^{\beta L_{\ell}}\right] = \sum_{j=1}^{|\mc V_{\ell}|} e^{\beta\lambda_{j}(L_{\ell})} > 0,
\]
where $\lambda_{j}(L_{\ell})$ are the real non-negative eigenvalues of the symmetric positive semi-definite matrix $L_{\ell}$. Thus $0 \leq O_{\ell}(\beta) < E_{\ell}(\beta)$. Dividing by $E_{\ell}(\beta)>0$ proves the result.
\end{proof}
In terms of $R_{\ell}(\beta)$, the average sign in Eq.~\eqref{eq:average-sign-bipartivity} and generalized bipartivity in Eq.~\eqref{eq:generalized-bipartivity} take the forms
\be\label{eq:s_B_in_terms_of_r}
\langle s\rangle_{\ell,\beta} = \frac{1-R_{\ell}(\beta)}{1+R_{\ell}(\beta)}, \qquad \mc{B}_{\ell}(\beta) = \frac{1}{1+R_{\ell}(\beta)}.
\ee
Their bounds follow immediately:
\be\label{eq:bipartivity-sign-bounds}
\frac12<\mc B_\ell(\beta)\leq1, \qquad 0<\langle s\rangle_{\ell,\beta}\leq1.
\ee
The upper bound is reached when $O_\ell(\beta)=0$, so all
contributing closed sequences have an even number of hops.
As the odd-sector weight approaches the even-sector weight,
$\mc B_\ell(\beta)$ approaches $1/2$ and the average sign
approaches zero. The same dynamics is responsible for the bounds in Estrada bipartivity~\eqref{eq:beta_range}.

For $\beta>0$, the upper bound characterizes bipartiteness of the HG.
If the HG is bipartite, every closed walk contains an even number
of hops. Diagonal stays preserve this parity, so $O_\ell(\beta)=0$
and the average sign is unity at every temperature. Conversely, if the HG contains an odd cycle of length $g$, then
$\operatorname{Tr}(W_\ell^g)>0$. The corresponding
term
\[
\frac{\beta^g}{g!} \operatorname{Tr}(W_\ell^g)
\]
makes a strictly positive contribution to $O_\ell(\beta)$ for
every $\beta>0$. Therefore,
\be\label{eq:bipartivity-equality}
\mathcal B_\ell(\beta)=1 \quad\Longleftrightarrow\quad \Gamma_\ell(G)\text{ is bipartite},\quad \forall\beta>0.
\ee
Thus, the average sign being unity also implies the HG is
bipartite. The restriction $\beta>0$ is essential: at $\beta=0$,
only the empty word contributes, giving $\mathcal B_\ell(0)=1$
for every HG.
In the high-temperature limit, $\beta\to0$ and since $E_\ell(0)=|\mc V_\ell|$ and $O_\ell(0)=0$,
\[
\lim_{\beta\to0}\mathcal B_\ell(\beta)=1, \qquad \lim_{\beta\to0}\langle s\rangle_{\ell,\beta}=1.
\]
This limit holds for every HG, including nonbipartite graphs: the factors $\beta^n \to 0$ suppress all contributions containing hops.

To determine the leading order term in the expansion of average sign in terms of $\beta$, we recall that $W_\ell$ is off-diagonal. A sequence containing only one hop cannot close, regardless of how many diagonal stays it contains. The first possible odd contribution therefore comes from three-hop closed walks:
\[
O_\ell(\beta) = \frac{\beta^3}{6} \operatorname{Tr}(W_\ell^3) +\mc O(\beta^4).
\]
Using $E_\ell(\beta)=|\mc V_\ell|+\mathcal O(\beta)$, we obtain
\be\label{eq:bipartivity-high-temperature}
\mc B_\ell(\beta) = 1-\frac{\operatorname{Tr}(W_\ell^3)}{6|\mc V_\ell|}\beta^3
+\mc O(\beta^4).
\ee
The average sign then becomes,
\be\label{eq:average-sign-high-temperature}
\langle s\rangle_{\ell,\beta} = 1-\frac{\beta^3}{3|\mc V_\ell|}\operatorname{Tr}\left(W_{\ell}^{3}\right) + \mc O(\beta^4).
\ee
The trace in Eq.~\eqref{eq:average-sign-high-temperature} sums the weights of closed walks of three
hops on the HG. Each such walk traverses a triangle, giving
\be\label{eq:weighted_triangle_trace}
\operatorname{Tr}(W_\ell^3) = 6\sum_{\{ \sigma,\tau,\rho\}\in\mc T_\ell} (W_\ell)_{\sigma\tau}(W_\ell)_{\tau\rho}(W_\ell)_{\rho\sigma}.
\ee
Here $\mc T_\ell$ is the set of HG triangles, each counted once, with $|\mc T_\ell|=N_{\triangle}(\Gamma_\ell)$ as defined in Eq.~\eqref{eq:triangle}. Recalling $W_\ell = \frac{J_1}{2}A_{\ell,1}+\frac{J_2}{2}A_{\ell,2}$ from Eq.~\eqref{eq:full_weighted_adj_matrix}, each matrix element contributes $J_r/2$ for a range-$r$ edge. Thus, Eq.~\eqref{eq:weighted_triangle_trace} weights each triangle by the product of its three edge weights. The factor of six accounts for three starting vertices and two traversal directions. Thus, the leading possible reduction of the average sign is determined
by the total weighted triangle contribution per HG vertex.
If the HG contains no triangles, this cubic term vanishes.

To calculate the higher-order corrections, we use the odd-to-even ratio $R_\ell (\beta)$ defined in Theorem~\ref{thm:odd-to-even-walk}.
\[
E_{\ell}(\beta) = \sum_{n=0}^{\infty} e_{n}\beta^{n}, \,\, O_{\ell}(\beta) = \sum_{n=0}^{\infty} o_{n}\beta^{n}, \,\, R_\ell (\beta) = \sum_{n=0}^{\infty} r_{n}\beta^{n}.
\]
Here, $e_n$ and $o_n$ collect the traces of operator words with an even or odd number of $W_\ell$ factors, respectively, including the factor $1/n!$. Substituting these series into the identity $O_\ell(\beta) = R_\ell (\beta) E_{\ell}(\beta)$ and equating coefficients of the same power of $\beta$ gives
\[
o_n = \sum_{k=0}^{n}e_k r_{n-k}.
\]
Since $e_0 = |\mc V_\ell|$, the coefficients can be determined recursively:
\[
r_0 = \frac{o_0}{|\mc V_\ell|}, \qquad r_n = \frac{1}{|\mc V_\ell|}\left(o_n - \sum_{k=1}^{n}e_k r_{n-k} \right), \qquad n \geq 1
\]
For the loopless HG, $o_0 = o_1 = o_2 = 0$, and hence $r_0 = r_1 = r_2 = 0$. Through fifth order, the ratio therefore takes the form
\[
R_\ell (\beta) = r_3\beta^3+r_4\beta^4+r_5\beta^5 +\mc O(\beta^6)
\]
Using the exact relation between this ratio and the average sign,
\[
\langle s \rangle_{\ell,\beta} = \frac{1-R_\ell (\beta)}{1+R_\ell (\beta)} = 1-2R_\ell (\beta) + 2R_\ell (\beta)^{2} + \mc O(R_\ell (\beta)^{3})
\]
we obtain
\[
\langle s \rangle_{\ell,\beta} = 1 - 2r_3 \beta^3 - 2r_4 \beta^4 - 2r_5 \beta^5 + \mc O(\beta^6),
\]
where
\[
\begin{aligned}
r_4 &= \frac{\operatorname{Tr}(D_\ell W_\ell^3)}
{6|\mc V_\ell|} - \frac{\operatorname{Tr}(D_\ell)
\operatorname{Tr}(W_\ell^3)}{6|\mc V_\ell|^2}, \\
r_5 &= \frac{\operatorname{Tr}(D_\ell^2W_\ell^3)+\operatorname{Tr}(D_\ell W_\ell D_\ell W_\ell^2)}{24|\mc V_\ell|}+\frac{\operatorname{Tr}(W_\ell^5)}{120|\mc V_\ell|}
\\[1mm]
&-\frac{\operatorname{Tr}(D_\ell)\operatorname{Tr}(D_\ell W_\ell^3)}
{6|\mc V_\ell|^2}+\frac{[\operatorname{Tr}(D_\ell)]^2
\operatorname{Tr}(W_\ell^3)}{6|\mc V_\ell|^3}
\\[1mm]
&-\frac{\left[\operatorname{Tr}(D_\ell^2)+\operatorname{Tr}(W_\ell^2)
\right]\operatorname{Tr}(W_\ell^3)}{12|\mc V_\ell|^2}.
\end{aligned}
\]
Fig.~\ref{fig:odd_walk_HG}~shows a diagrammatic representation of odd closed walks contributing to the ratio $R_\ell(\beta)$.
In Ref.~\cite{Hen2021}, Hen relates QMC configuration signs to closed-walk phases on the Hamiltonian graph. Here, we connect the average sign to generalized Estrada bipartivity while retaining diagonal contributions.

The higher-order corrections contain both longer odd closed walks and diagonal stays along shorter walks. At fourth order, the odd sector contains triangular walks with one stay. At fifth order, it contains five-hop closed walks and triangular walks with two stays. The trace $\mathrm{Tr}\left(W^5\right)$ includes simple five-cycles as well as walks that traverse a triangle and retrace an edge. The mixed traces involving $D_\ell$ weight the triangular walks according to the vertices at which the stays occur.

Each diagonal stay increases the order in $\beta$ without changing the hopping parity. Normalization by $E_\ell(\beta)$ also introduces products of traces, such as $\mathrm{Tr}\left(W^2\right)\mathrm{Tr}\left(W^3\right)$. Although such products do not represent a single closed walk, they retain an odd total number of $W_\ell$ factors. Thus, the coefficients $r_4$ and $r_5$ incorporate both the odd-walk weights and their normalization by the even-sector weight.

Finally, the generalized bipartivity reduces to Estrada's
adjacency-based expression when the diagonal part of the
Hamiltonian is constant. For the HAF, this occurs when the HG
has constant weighted degree, $D_\ell=d_\ell I$. In this case,
\[
    \operatorname{Tr}
    e^{\beta(D_\ell\pm W_\ell)}
    =
    e^{\beta d_\ell}
    \operatorname{Tr}e^{\pm\beta W_\ell}.
\]
The common factor cancels in the average sign, giving
\[
    \langle s\rangle_{\ell,\beta}
    =
    \frac{\operatorname{Tr}e^{-\beta W_\ell}}
         {\operatorname{Tr}e^{\beta W_\ell}}.
\]
Using $\mathcal B_\ell(\beta)
=\tfrac12(1+\langle s\rangle_{\ell,\beta})$, we obtain
\[
\mathcal B_\ell(\beta)
=
\frac{
    \operatorname{Tr}
    \bigl(e^{\beta W_\ell}+e^{-\beta W_\ell}\bigr)
}{
    2\,\operatorname{Tr}e^{\beta W_\ell}
}
=
\frac{
    \operatorname{Tr}\cosh(\beta W_\ell)
}{
    \operatorname{Tr}e^{\beta W_\ell}
}.
\label{eq:estrada-bipartivity}
\]
This is Estrada's spectral bipartivity evaluated on the scaled
adjacency matrix $\beta W_\ell$: the numerator collects the
weights of even closed walks, while the denominator collects
those of all closed walks. The same reduction applies when
the diagonal part of the Hamiltonian vanishes.

To summarize, the average sign is fixed exactly by the balance between even and odd closed walks on the HG. It equals unity if and only if the HG is bipartite, it first departs from unity at the order set by the shortest odd cycle, and the vertex degrees enter one order later through their correlation with the local triangle weight.

\section{Conclusion \& Outlook}\label{sec:conclusion}
In this work, we studied the GS of the \JJ HAF Hamiltonian as an optimization problem on the HG, the weighted graph of computational-basis configurations generated by the Hamiltonian's off-diagonal exchange terms in a fixed-magnetization sector. The HG is the token graph of the underlying lattice, and the identity $H_\ell = E_F - L_\ell$ turns the HAF eigenvalue problem into a spectral problem on it, so that computing the GS energy is equivalent to solving the spectrally relaxed \mct on the HG. At fixed GS amplitudes and with binary phases, the phase-dependent part of the variational energy is a classical antiferromagnetic Ising model on the HG with amplitude-dressed couplings, equivalent to weighted \mct on the same HG. 

For weights generated by the exact GS amplitudes, the global minimum of
the XY energy lies at a $\{0,\pi\}$ assignment, since the GS of the real
Hamiltonian can be chosen real, and both the \rt and GW relaxations
recover the GS signs with high accuracy. For a nondegenerate GS, the GW
relaxation is exact: its minimum equals $E_0$, its optimum is the
rank-one matrix built from the GS signs, and hyperplane rounding
recovers these signs in a time polynomial in the size of the HG.

In VMC with separate amplitude and phase parameters, each phase update
is a \rt step for weighted \mct on the HG with weights set by the
current amplitudes, so that VMC acts as an evolving \rt relaxed
\mct and the phase ansatz inherits its spurious minima and saddle
points, consistent with the rugged landscape reported for NQS. Because
the MSR solves this \mct for any amplitudes on a bipartite HG, the
optimal phases do not move during training, which offers a possible explanation
for why sign learning is easier on bipartite lattices. With a real-valued
ansatz, a sign prior fixes a cut of the HG in advance, and a good GS
energy requires this cut to agree with the GS cut on the high-weight
edges, so the signs of the configurations with the largest amplitudes
matter most.

On a bipartite HG, the optimal phases come from its two-coloring, which reproduces the MSR, while the odd cycles of the HG, inherited from those of the lattice, prevent any such coloring. The HG edges carry the off-diagonal elements of $H_\ell$ as weights; this indicates that a diagonal unitary causes the sector Hamiltonian to be stoquastic if and only if the HG is bipartite, the residual non-stoquasticity constitutes a \mct{} deficit, and the average sign of the QMC gives the generalized Estrada bipartivity of the HG.

Our results show that the HG viewpoint reveals physics that
is hard to see in the lattice formulation of the HAF: the MSR appears as
the two-coloring of a bipartite HG, and the sign learning obstruction of frustrated
HAFs as odd cycles of the HG and a \mct deficit. This suggests that the
same viewpoint can serve as a framework for other strongly correlated
systems. A natural next step is the Fermi-Hubbard model. For fixed
numbers of up and down electrons, its HG is the Cartesian product of two
HGs of the type studied here, one for each spin species, but the hopping
amplitudes can carry fermionic signs and the on-site repulsion is not
related to the vertex degree, so the identity $H_\ell = E_F - L_\ell$
does not carry over directly. Since the HAF is the strong-coupling limit
of the half-filled Hubbard model, the HG studied here is expected to
emerge from the Hubbard HG once doubly occupied configurations are
eliminated, which would connect the sign structures of the two models.

\medskip
\textsf{Code Availability.}
The code used to generate the results for this paper is available in the GitHub repo in Ref.~\cite{shamim2026code} and as ancillary files on arXiv. NetKet~\cite{netket3:2022} was
used in this work.

\smallskip
\textsf{Acknowledgments.} We are grateful to Prof. Zohar Nussinov, Prof. Marin Bukov, Prof. Ernesto Estrada, Prof. Andrey Bagrov, Prof. Markus Schmitt, Prof. Talal Ahmed Chowdhury, Prof. Andrew Lucas, Prof. Bojan Mohar, Prof. Sam Hopkins, Prof. Filippo Vicentini, Prof. David A. Huse and Dr. Tao Liu for insightful discussions and comments. We thank the anonymous reviewer for a detailed report and constructive comments, which helped to improve the manuscript. M.H. acknowledges support from the Natural Sciences and Engineering Research Council of Canada (NSERC).

\bibliographystyle{apsrev4-2}
\bibliography{ref}

\end{document}